\documentclass[11pt,cite,epsfig,psfrag]{article}
\usepackage[utf8]{inputenc}
\usepackage{placeins}
\usepackage{geometry}
 \usepackage{graphicx}
 \usepackage{subfig}
  \usepackage{amsmath} 
  \graphicspath{{images_massive_Rup_plots/}}
  \usepackage[usenames, dvipsnames]{color}
\usepackage{wrapfig}
\usepackage{boxedminipage}

\usepackage{fullpage}
\usepackage{amsmath}
\usepackage{amssymb}
\usepackage{graphicx}
\usepackage{mathrsfs}
\usepackage{wrapfig}
\usepackage{boxedminipage}
\usepackage{epsfig}
\usepackage{setspace}
\usepackage{float}
\usepackage{hyperref}
\usepackage{enumerate}
\usepackage{cite}
\usepackage[font=footnotesize,labelfont=rm]{caption}

\usepackage[numbers,sort&compress]{natbib}

\def\be{\begin{equation}}
\def\ee{\end{equation}}
\def\bea{\begin{eqnarray}}
\def\eea{\end{eqnarray}}

\numberwithin{equation}{section}

 \newcommand{\RN}[1]{%
   \textup{\uppercase\expandafter{\romannumeral#1}}%
 }
\begin{document}

\thispagestyle{empty}

\vskip 2cm

\begin{center}
	{\Large \bf Ruppeiner Geometry, Phase Transitions and Microstructures of Black Holes in Massive Gravity}
\end{center}

\vskip .2cm

\vskip 1.2cm

\centerline{  \bf  Pavan Kumar Yerra \footnote{pk11@iitbbs.ac.in}  and Chandrasekhar Bhamidipati \footnote{chandrasekhar@iitbbs.ac.in}
}

\vskip 7mm 
\begin{center}{ School of Basic Sciences\\ 
Indian Institute of Technology Bhubaneswar \\ Bhubaneswar 752050, India}
\end{center}

\vskip 1.2cm
\vskip 1.2cm
\centerline{\bf Abstract}
Using the new normalized thermodynamic scalar curvature, we investigate the microstructures and phase transitions of black holes in massive gravity for horizons of various topologies. We find that the graviton mass enhances the repulsive interactions of small black holes and weakens the attractive interactions of large black holes, with possiblity of new repulsive regions for microstructures in phase space.  In addition, the repulsive interactions of small black hole are strong for spherical topology, followed by flat and hyperbolic topology; while, the attractive interactions of large black hole are strong for hyperbolic topology, followed by flat and weakest for spherical topology.\\



\newpage
\setcounter{footnote}{0}
\noindent

\baselineskip 15pt

\section{Introduction}

\noindent

Geometrical methods have long been applied to the study of thermodynamics and demonstrated their usefulness in providing deep insights in to various aspects of black holes. Starting from a metric constructed out of a suitable thermodynamic potential, with phase space composed of other thermodynamic variables, scalar curvature $R$ of the state space encodes the information about phase transitions and critical points. Today, there are several competing methods to set up a thermodynamic geometry depending on the potential chosen. For instance, in the Weinhold\cite{Weinhold} and Ruppeiner's\cite{Ruppeiner} methods, internal energy and entropy respectively, are chosen as thermodynamic potentials, based on which the metric of the state space is constructed (see also~\cite{Quevedo:2007mj,Hendi:2015rja} for other variations~\cite{Mansoori1}). The set up of thermodynamic geometry has now been applied to a broad class of systems such as ideal and van der Waals fluids, Ising models, quantum Bose and Fermi gas etc. (see \cite{Ruppeiner:2012uc,RuppeinerRMP} and references there in). The understanding developed from the aforementioned explorations is that the line element of the thermodynamic geometry is a measure of the distance between two near by fluctuations of the system, with the distance going to infinity if the fluctuations between neighboring states are less probable. For example, R is zero for the classical ideal gas, where as it is positive (negative) for Fermi (Bose) gases. The sign of R can then be used to know about the nature of microstructures of the thermodynamic system: A positive or a negative R signifies the presence of repulsive or attractive interactions, respectively among the microscopic degrees of freedom, where as a vanishing R means a non-interacting system. Thus, the methods of thermodynamic geometry could be used as macroscopic probes to understand the microscopic interactions of the degrees of freedom. \\

\noindent
Particularly, a macroscopic interpretation of thermodynamics of black holes exists for long time \cite{Hawking:1974sw,Hawking:1976de,Bekenstein:1973ur,Bekenstein:1974ax,Bardeen:1973gs,Gibbons:1976ue}, where thermodynamic quantities can be obtained directly from an action principle, but, the microscopic derivation is not available yet, except for certain thermodynamic quantities in special situations~\cite{Strominger:1996sh,Rovelli:1996dv,Sen:2007qy}.  A major line of research in this direction is to understand the thermodynamics and phase transitions of black holes in AdS~\cite{Hawking:1982dh}, which have a very nice connection with the van der Waals liquid-gas system~\cite{Chamblin:1999tk,Chamblin:1999hg}  in extended thermodynamic phase space approach~\cite{Kubiznak:2012wp}, based on a dynamical cosmological constant scenario~ \cite{Henneaux:1984ji,Teitelboim:1985dp,Henneaux:1989zc,Caldarelli:1999xj,Padmanabhan:2002sha,Wang:2006eb,Sekiwa:2006qj,LarranagaRubio:2007ut,Kastor:2009wy,Dolan:2010ha,Cvetic:2010jb,Kubiznak:2012wp,Gunasekaran:2012dq,K,Kubiznak:2016qmn,majhi}. Thus, the thermodynamic geometry approach is quite useful considering the present status of thermodynamics of black holes, where, the traditional method of understanding the system from a microscopic statistical mechanical point of view is not fully developed yet. Indeed, R though being a macroscopic quantity, can be used to probe microscopic structure of thermodynamic systems, such as, black holes. R has been calculated now for a wide variety of black hole systems, starting from initial calculations for BTZ black holes in~\cite{Cai} to more works in last few years (see~\cite{Shen,cai1,cai2,Ruppeiner2,Ruppeiner3,Janyszek:1989zz,Janyszek1990,Wei2015,Dolan:2015xta,AR,Deng,Deng1,Chabab,Dehyadegari,Mansoori1,Banerjee:2010da,Sahay:2010tx} for a partial list). Recently, a novel calculation of R using a new normalized curvature and the methods of extended thermodynamic phase space, have thrown interesting connections and dissimilarities for the first time between van der Waals fluid system and black holes~\cite{Wei:2019uqg,Wei:2019yvs}. Particularly, in the context of charged black holes in AdS, it was shown that although an attractive type interaction dominates for most of the parameter space of the system, there is still a weakly repulsive behavior for the small black holes at high temperatures. This behavior of the black hole system is in contrast to the behavior in van der Waals systems, which only have an attractive type interaction among their microscopic molecules~\cite{Wei:2019uqg,Wei:2019yvs}. Subsequently, studying the microstructures of black holes and the nature of their interactions has been actively pursued, with improved understanding for several other systems~\cite{Xu:2019gqm,Wei:2019ctz,Ghosh:2019pwy,Naveen,Wei:2020poh,Mahish:2020gwg,Xu:2019nnp,Xu:2020gzm,Xu:2020ftx,Kumara:2020ucr,Kumara:2019xgt}. Our aim here is to extend the above works to see the effect of introducing graviton mass and presence of horizons of more general topologies on the thermodynamics, phase transitions and microstructures of black holes. It is certainly an interesting question to check whether the Ruppeiner scalar can effectively change in systems with different horizon topologies, especially, because the massive gravity system was an important example where phase transitions were first noted in non-spherical topology situations~\cite{Cai:2014znn,Hendi:2017fxp}.\\

\noindent

Furthermore, General theory of relativity has made important predictions which have received success with experimental 
confirmations, more interestingly, with nice agreements with recent
observations of LIGO collaboration~\cite{LIGO2017,deRham2014Review} pertaining to gravitational waves. There are however many other phenomena, such as, the accelerated expansion of our universe, the hierarchy problem, the cosmological constant issue, to name a few, which have spurred interest in looking for models beyond Einstein's theory. In this regard, one interesting direction of research is to consider massive graviton theories, keeping the hierarchy problem and quantum gravity in mind~\cite{MassiveIb,MassiveIc}, which seems have good input from recent data~\cite{Abbott}, generating lower bounds on the mass of gravitons. Models with introduction of graviton's mass have a rather long history, with initial studies done by Fierz and Paullo in 1939~\cite{Fierz1939}, which has now undergone many changes with the inclusion of novel ideas, e.g., new massive gravites\cite{BDghost,Newmasssive,dRGTI,dRGTII} (actively pursued in literature \cite{Cai:2014znn,NewM1,NewM2,NewM3,NewM4,NewM5,HassanI,HassanII,EslamPanah:2019fci,EslamPanah:2018rob,EslamPanah:2018evk}). More over, there has been a continued interest in studying the thermodynamical properties of black holes in these systems\cite{BHMassiveI,BHMassiveII,BHMassiveIII,BHMassiveIV}, with applications to astrophysical and cosmological situations, particularly to capture deviations from Einstein's gravity\cite{Katsuragawa,Saridakis,YFCai,Leon,Hinterbichler,Fasiello,Bamba}. We should point out that there are also models of massive gravity with  holographic motivations in mind~\cite{Vegh}, which make use of metric which is singular, pointing towards the fact that graviton mass might give rise to stable ghost-free gravity\cite{HZhang}, including situations which involve black holes\cite{PVMassI,PVMassII,PVMassIII,Cai:2014znn,Hendi:2017fxp,PVMassV}. Furthermore, apart from applications which involve addressing issues in Einstein's gravity\cite{Gumrukcuoglu,Gratia,Kobayash,DeffayetI,DeffayetII,DvaliI,
DvaliII,Will,Mohseni,GumrukcuogluII,NeutronMass,Ruffini}, there is also progress in using massive gravity theories to explain current experimental observations of dark matter\cite{Schmidt-May2016DarkMatter} and explaining accelerated expansion of our universe without making use of dark energy component\cite{MassiveCosmology2013,MassiveCosmology2015}. It is worth mentioning that there have been various 
works in trying to embed models of massive gravities in schemes of string theory~\cite{MGinString2018}. What is more important for the work pursued in the present paper is that, a van der Waals type liquid gas phase transition in the extended thermodynamic phase space is confirmed to exist and also actively being studied for new physics, beyond the standard charged black holes in AdS~\cite{PVMassV,Hendi:2017fxp,Alberte,Zhou,Dehyadegari,Magmass,Hennigar:2017apu,Hendi:2017bys,Yerra:2020bfx}.  Ruppeiner geometry in dRGT massive gravity was studied in~\cite{Chabab:2019mlu}, but the issue of microstructures were not addressed, which is what we pursue in this work.\\

\noindent
The paper is organized as follows. In section-(\ref{massiveIntro}), we collect main formulas on thermodynamics of charged black holes in massive gravity theories with a short summary of its phase structure. Section-(\ref{Rmassive}) contains our main results on Ruppeiner geometry and its analysis in various cases. Here we present how thermodynamic scalar curvature varies in the massive gravity theories with its parameters, and also its variation with topology of horizon. In section-(\ref{conclude}), we end with a summary of results and conclusions.

\section{AdS Topological Charged  Black Holes in Massive Gravity} \label{massiveIntro}
We employ the action for Einstein-Maxwell theory in massive gravity with a negative cosmological constant $\Lambda$ in 4-dimensions as ~\cite{Cai:2014znn,Hendi:2017fxp,Hendi:2017bys}:

\begin{equation}
I=\frac{-1}{16\pi }\int d^{4}x\sqrt{-g}\left( \mathcal{R}-2\Lambda
-\mathcal{F}+m^{2}\sum_{i}^{4}c_{i} \, \mathcal{U}_{i}(g,f)\right) ,
\end{equation}
where  $\mathcal{R}$ is the Ricci scalar, and the Maxwell invariant is $\mathcal{F}=F_{\mu \nu}F^{\mu \nu }$ with  $F_{\mu \nu }=\partial _{\mu
}A_{\nu }-\partial _{\nu }A_{\mu }$ given in terms of gauge
potential $A_{\mu }$, and  $m$ is graviton mass term.  Other parameters appearing in the action, such as, $c_{i}$'s are   constants. Further, the $\mathcal{U}_{i}$'s
are symmetric polynomials constructed out of the eigenvalues of the $4\times 4$ matrix $ \mathcal{K}_{\nu }^{\mu }=\sqrt{g^{\mu \alpha }f_{\alpha \nu }}$, which is known to be:
\begin{eqnarray}
	\mathcal{U}_{1}&=&\left[ \mathcal{K}\right], \nonumber \\ \mathcal{%
		U}_{2} &= & \left[ \mathcal{K}\right] ^{2}-\left[ \mathcal{K}^{2}\right] , \nonumber \\ 
	\mathcal{U}_{3}&=&\left[ \mathcal{K}\right] ^{3}-3\left[ \mathcal{K}\right] %
	\left[ \mathcal{K}^{2}\right] +2\left[ \mathcal{K}^{3}\right], \nonumber \\ 
	\mathcal{U}_{4}&=&\left[ \mathcal{K}\right] ^{4}-6\left[ \mathcal{K}^{2}%
	\right] \left[ \mathcal{K}\right] ^{2}+8\left[ \mathcal{K}^{3}\right] \left[
	\mathcal{K}\right] +3\left[ \mathcal{K}^{2}\right] ^{2}-6\left[ \mathcal{K}%
	^{4}\right].
\end{eqnarray}
and $f$ is a reference metric. With the assumption of a static ansatz, the equations of motion following from the above action are known to admit black holes solutions with various horizon topologies, denoted by the metric~\cite{Cai:2014znn,Hendi:2017fxp,Hendi:2017bys}:
\begin{equation}
ds^{2}=-Y(r) dt^{2}+\frac{dr^{2}}{Y(r) } + r^{2}h_{ij}dx_{i}dx_{j} \ ,
  \label{eq:metric}
\end{equation}
where the reference metric $f_{\mu \nu}$ is:
\begin{equation} f_{\mu \nu
}=\text{diag}(0, 0, c_0^{2} h_{ij}) \, .
\end{equation}
Here, $c_0$ is a positive constant and $i,j$ run over the indices $1,2$. $h_{ij}dx_{i}dx_{j}$ stands for a spatial metric of
constant curvature $2k$ having volume $4\pi$. $ k $ can actually take various values, such as, +1, 0, or -1, corresponding respectively to  a spherical, flat, or hyperbolic topology of the black hole horizon. Using the form of $f_{\mu \nu}$, the $\mathcal{U}_{i}$'s can be shown to be~\cite{Cai:2014znn,Hendi:2017fxp,Mo:2017nes}:
\begin{equation}
\mathcal{U}_{1}= \frac{2c_0}{r}, \quad \mathcal{%
	U}_{2} =  \frac{2c_0^2}{r^2} , \quad
\mathcal{U}_{3}= 0 , \quad
\mathcal{U}_{4}= 0 \ ,
\end{equation}
where one sets,  $c_3 =c_4 = 0$, since $\mathcal{U}_{3}= \mathcal{U}_{4}= 0$.  Using an ansatz for the gauge potential $A_{\mu }=h(r)\delta
_{\mu }^{0}$, the lapse function $Y(r)$  is given by~\cite{Cai:2014znn,Hendi:2017fxp,Hendi:2017bys}:
\begin{equation}
Y(r) =k-\frac{m_{0}}{r}-\frac{\Lambda r^{2}}{3}+\frac{q^{2}}{ r^{2}}+m^{2}(\frac{c_0c_{1}}{2}r+c_0^{2}c_{2}) \, .
 \label{Y(r)}
\end{equation}
Here, the integration constants  $m_0$ and $q$ are related to the mass $M$ and the electric charge $Q$ of the hole, respectively. It is useful to remember that the solution~\eqref{Y(r)}, is asymptotically AdS and in the absence of graviton mass $(m =0)$, it goes back to the standard Reissner-Nordstrom black hole solution~\cite{Hendi:2017fxp}. It should also be mentioned that the choice of the reference metric makes the additional massive gravity terms to give rise to Lorentz-breaking property~\cite{Vegh}. 

\vskip 0.5cm
\noindent
Using the largest positive root $r_+$ of $Y(r_+)=0$, various thermodynamic quantities of the system, such as, the temperature $T$,  mass $M$ and electric potential $\Phi$ can be expressed as~\cite{Hendi:2017fxp}: 
\begin{eqnarray}
T &=&\frac{k}{4\pi r_{+}}-\frac{r_{+}\Lambda }{4\pi
}-\frac{q^{2}}{4\pi r_{+}^{3}}  +\frac{m^{2}}{4\pi r_{+}}\left(
c_0c_{1}r_{+}+c_{2}c_0^{2}\right) \ ,
\label{eq:Temp} \\
 M &=&\frac{m_{0}}{2 } = \frac{r_+}{2} \left(k -\frac{\Lambda}{3}r_+^2 + \frac{q^2}{r_+^2} + m^2(\frac{c_0 c_1}{2}r_+ + c_0^2c_2)\right)\ ,  \label{eq:Mass} \\
\Phi &=& A_{\mu }\chi ^{\mu }\left\vert _{r\rightarrow \infty }\right. -A_{\mu
}\chi ^{\mu }\left\vert _{r\rightarrow r_{+}}\right. =\frac{q}{r_{+}} \, ,
\end{eqnarray}
where  electric charge $Q=q$ (corresponding to the electric potential $\Phi$ of the black hole),
and entropy $S=\pi r_{+}^{2}$.
\noindent
In the extended thermodynamic phase space approach, one defines the pressure from a varying cosmological constant, using the relation $P=-\frac{\Lambda}{8\pi}$ with its conjugate taking the meaning of a thermodynamic volume $V$, which for static black holes turns out to be identical to geometric volume, as we see below. In this set up,  one usually identifies  the mass $M$ of the black hole as the enthalpy $H$~\cite{Kastor:2009wy}, which together with above thermodynamic quantities, satisfies the first law of 
black hole thermodynamics given as~\cite{Hendi:2017fxp}:
\begin{equation}
dM=TdS+\Phi dQ+VdP+\mathcal{C}_{1}dc_{1} \ ,  \label{1stlaw}
\end{equation}
where 
\begin{eqnarray}
V &=&\left( \frac{\partial M}{\partial P}\right) _{S,Q,c_{1}}=\frac{4\pi}{3} r_+^3 \ ,  \label{Vol} \\
\mathcal{C}_{1} &=&\left( \frac{\partial M}{\partial c_{1}}\right)
_{S,Q,P}=\frac{c_0m^{2}r_{+}^{2}}{4 } \ .
\label{C1} 
\end{eqnarray}
\noindent
Now, the  heat capacities of the black holes in massive gravity theory at constant volume $C_V$, and at constant pressure $C_P$ can be shown to be~\cite{Hendi:2017bys,Mo:2017nes}:  
\begin{equation}
	C_V=0 \ ; \, \, \,   C_P= 2S\Bigg(\frac{8PS^2+S(k+m^2c_0^2 c_2) -\pi q^2 + \frac{m^2c_0 c_1 S^{3/2}}{\sqrt{\pi}}}{8PS^2 -S(k+m^2c_0^2 c_2) + 3\pi q^2 } \Bigg) \ .
\end{equation}

\subsection{Phase Structure}
 Using $P=-\frac{\Lambda}{8\pi}$ in equation~\eqref{eq:Temp}, one gets the expression for equation of state $P(V, T)$ as: 
\begin{eqnarray} \label{eos1}
P & = & \frac{1}{8\pi}\Bigg\{\frac{(4\pi T-m^2c_0c_1)}{(\frac{3V}{4\pi})^{\frac{1}{3}}}-\frac{(k+m^2c_2c_0^2)}{(\frac{3V}{4\pi})^{\frac{2}{3}}}+\frac{q^2}{(\frac{3V}{4\pi})^{\frac{4}{3}}}\Bigg\} \ . \label{eq of st} 
\end{eqnarray}
One can also express this equation of state in terms of specific volume $v$ using $v=2r_+$. The equation of state~\eqref{eq of st} shows the Van der Waals behavior for topological charged black holes in massive gravity with the critical point obtained from the condition  
 $ \partial P/\partial V=\partial^2 P /\partial V^2=0 $, given by~\cite{Hendi:2017fxp}:  
\begin{equation}
	\label{eq:critical pt}
	\quad P_{\rm c}=\frac{\epsilon^2}{96\pi q^2 }\ , \quad V_{\rm c}=\frac{8 \sqrt{6}\pi q^3}{\epsilon^{\frac{3}{2}}} \ ,   \quad T_{\rm c}= \frac{\epsilon^{\frac{3}{2}}}{ 3\sqrt{6} \pi q} + \frac{m^2c_1 c_0}{4\pi} \ , 
\end{equation}
where $ r_{\rm c}= \sqrt{\frac{6}{\epsilon}}q \, $  and $\epsilon = (k +m^2c_2 c_0^2) > 0$. Here, we note that in terms of reduced parameters: $\tilde{P} = \frac{P}{P_c}, \quad \tilde{V} = \frac{V}{V_c}, \, \text{and} \, \, \tilde{T} = \frac{T}{T_c},  $ the equation of state is still depends on the parameters charge $q$ and graviton mass $m$. However, in a special case where the massive coefficient $c_1 = 0$,
the reduced equation of state is independent of charge and graviton mass, which is
\begin{equation}\label{red eq of st}
 \tilde{P} = \frac{8\tilde{T}}{3{\tilde{V}}^{1/3}} - \frac{2}{{\tilde{V}}^{2/3}} + \frac{1}{3{\tilde{V}}^{4/3}}.
\end{equation}
This reduced equation of state is exactly same as that of the 4-dimensional charged AdS  black hole and will give  the same phase structure in reduced parameter space given in~\cite{Wei:2019uqg,Wei:2019yvs}.
\begin{figure}[h!]
	
	{\centering
		\subfloat[]{\includegraphics[width=3.1in]{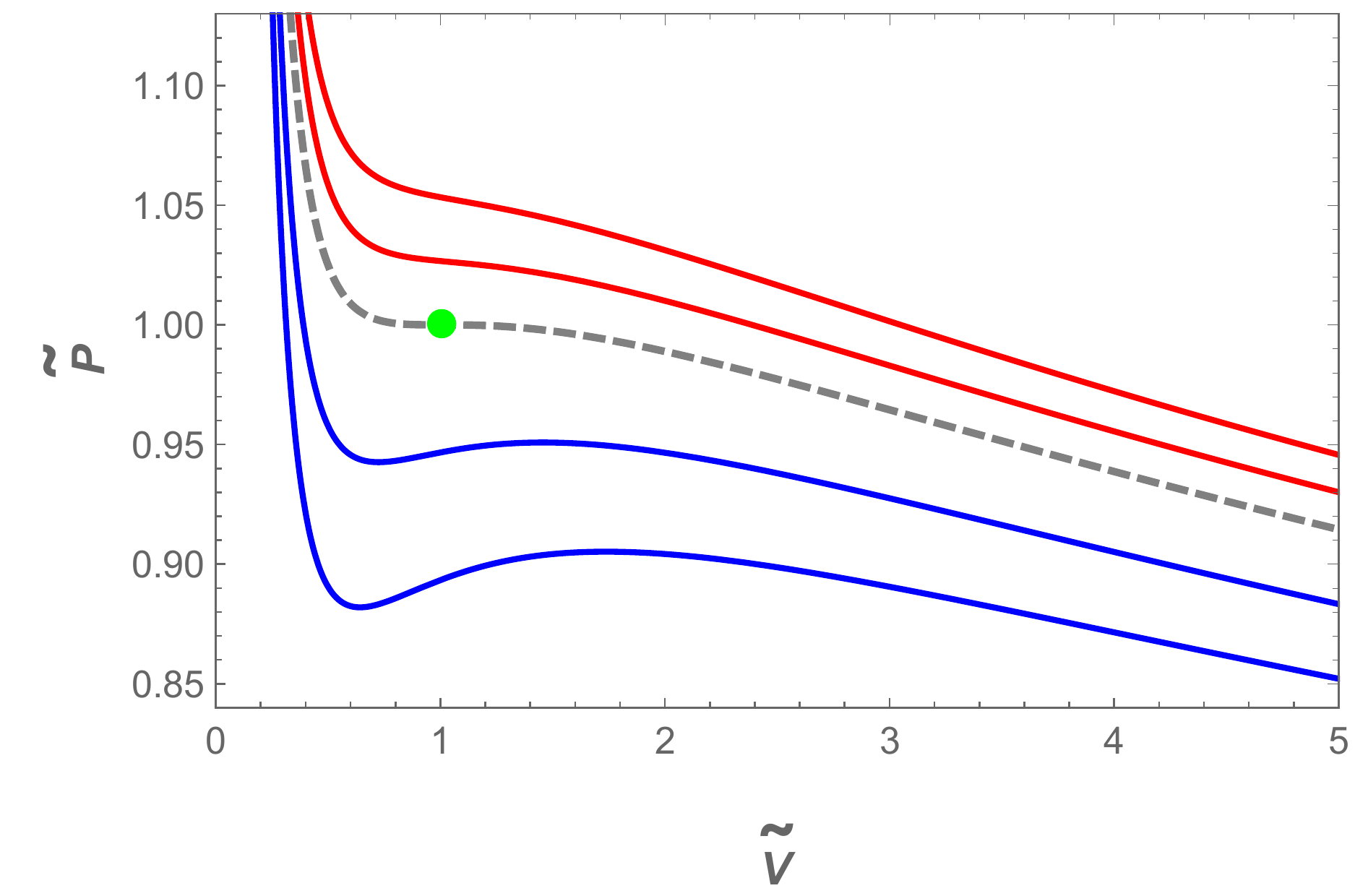}\label{fig:pv_plot}}		
		\subfloat[]{\includegraphics[width=3.1in]{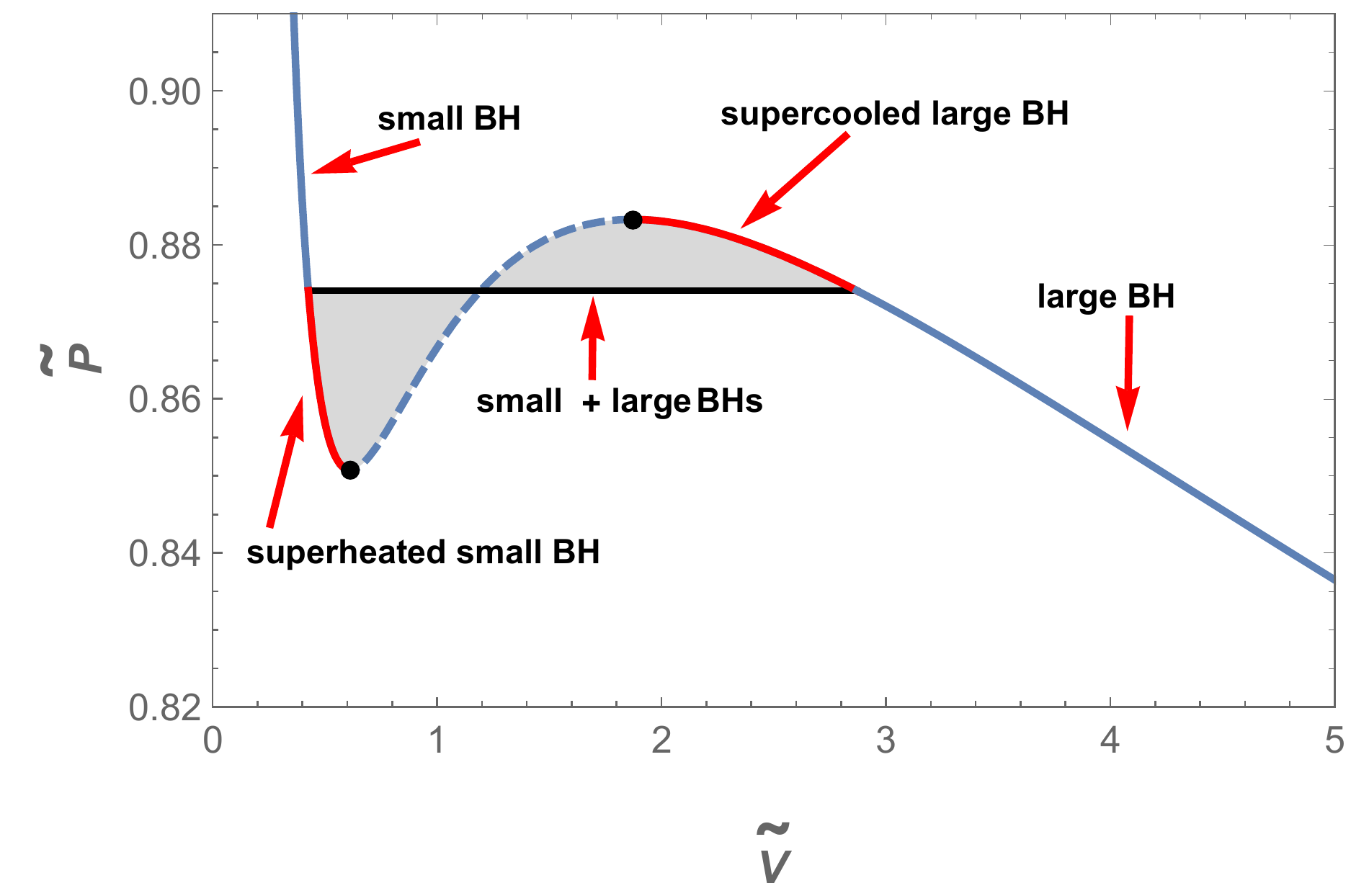}\label{fig:maxwell_plot}}

		\caption{\footnotesize (a) Sample isotherms for the reduced equation of state: Temperature $\tilde{T}$ of the isotherms decreases from top to bottom and the dashed isotherm is for $\tilde{T} =1$, while the critical point is shown in green colored dot. (b) A sample isotherm with $\tilde{T}<1$ under Maxwell's equal area law construction: The horizontal black colored line represents the phase transition pressure, the blue colored curves represent the stable phases of small and large black holes, the red colored curves represent the  metastable phases of superheated small black holes and supercooled large black holes, the dashed blue colored curve represents the unstable phase of black holes, and the black colored dots represent the spinodal points.  
		}   
	}
\end{figure}
\\ \noindent   We plot the isotherms for the reduced equation of state in Fig.~\eqref{fig:pv_plot}, where it shows  an ideal gas behavior indicating a unique phase of the black holes (supercritical black hole phase) for $\tilde{T} > 1$, while an oscillatory behavior for $\tilde{T} < 1$, indicating the existence of small and large black hole phases which undergo a first order phase transition that terminates at the critical point $(\tilde{V}, \tilde{P}) =(1,1)$. 
For an isotherm $\tilde{T} < 1$, the phase transition point can be obtained using Maxwell's equal area law~\cite{Spallucci:2013osa,Lan:2015bia}. As shown in Fig.~\eqref{fig:maxwell_plot}, the phase transition pressure which provides equal areas for an isotherm under Maxwell's construction, divides the isotherm into  small black hole phase (stable phase), large black hole phase (stable phase) and coexistence phase of small and large black holes. In a coexistence phase, there are metastable superheated small black hole and supercooled large black hole branches, and an unstable branch separated from metastable branches by spinodal points. 
\vskip 0.5cm
 \noindent The phase transition points, obtained using Maxwell's equal area law for the isotherms with $\tilde{T}<1$, form a curve called coexistence curve of small and large black holes, whose details are similar to the one done in~\cite{Spallucci:2013osa,Lan:2015bia,Wei:2019yvs}.
 Further, the spinodal points on a given isotherm with $\tilde{T}<1$, are the extremal points which form the small and large black hole spinodal curves, given by the condition~\cite{Wei:2019yvs},
\begin{equation}
(\partial_{\tilde{V} }\tilde{P})_{\tilde{T}}  =  0, \quad \text{or} \quad (\partial_{\tilde{V} }\tilde{T})_{\tilde{P}}  =  0.
\end{equation}
   In  $\tilde{T}-\tilde{V}$ plane, they are given by the following compact equation:
\begin{equation}
\tilde{T}_{\text{sp}} = \frac{3\tilde{V}^{\frac{2}{3}}- 1}{2\tilde{V}},
\end{equation}
where, $\frac{1}{3\sqrt{3}} < \tilde{V} < 1$ is for small black hole spinodal curve and $\tilde{V} > 1$ is for large black hole spinodal curve.
\vskip 0.5cm
\noindent
On the other hand, in the case of massive coefficient $c_1 \neq 0$, though we cannot move to reduced parameter space we proceed with non-reduced parameter space.  
In this case, one sees that the spinodal curve takes the form given by:
\begin{equation}\label{eq: spinodal curve c1 non zero}
T_\text{SP} =  \frac{ 2\epsilon(36\pi)^{1/3}V^{2/3}-16\pi q^2 + 3c_0 c_1 m^2V}{12\pi V}.
\end{equation}
However, we note that the expression for the coexistence curve is not readily available for this case. 
In what follows we study the Ruppeiner geometry of the system in the cases of massive coefficient $c_1 =0$ and $c_1 \neq 0$ in non-reduced space for consistency. 

\section{Microstructures of Black Holes} \label{Rmassive}

Empirically, it is well known that the scalar curvature $R$ of the thermodynamic geometry encodes information about whether  the interactions among the microstructures of the system are attractive ($R<0$) or repulsive ($R>0$). On the other hand $R=0$ gives rise to a sign changing curvature and $R=\infty$ captures the critical points of phase transitions or other physical aspects of thermodynamics (e.g., extremal limit etc.). Although,  there are discussions on the limitations of using scalar curvature in characterizing thermodynamic systems~\cite{Dolan:2015xta}, but the well known relations in black hole thermodynamics make the Ruppeiner geometry methods quite plausible, to  phenomenologically get a qualitative idea about the nature of interactions of microstructures.  This analysis can then be matched with results of well known statistical mechanical systems to gain a better understanding, especially, where a microscopic picture is still under development. Aforementioned features of thermodynamic geometry approach are also supported by fluctuation theory and statistical mechanical models, making it an indispensable tool with advantages over other methods. Furthermore, the recent results on the proposal of using a new diagonal thermodynamic state space metric and a novel normalized curvature scalar to probe charged black holes in AdS has already given exciting results~\cite{Wei:2019uqg,Wei:2019yvs,Wei:2019ctz} (see~\cite{Xu:2019gqm,Ghosh:2019pwy} for other related methods). The metric in the construction of ref~\cite{Wei:2019uqg,Wei:2019yvs} takes the form:
\begin{equation} \label{metricG}
 ds^2=\frac{1}{T} \left( -\left(\frac{\partial^2 \mathcal{H}}{\partial T^2}\right)_{V} dT^2+\left(\frac{\partial^2 \mathcal{H}}{\partial V^2}\right)_{T} dV^2 \right),
\end{equation}
with  the Helmholtz free energy given as $\mathcal{H}=H-PV-TS$. As seen, from eqn. (\ref{metricG}), the fluctuation coordinates are  temperature $T$ and thermodynamic volume $V$, which are both extensive quantities matching the proposals of Ruppeiner geometry~\cite{Ruppeiner}. It is also possible to have entropy $S$ and pressure $P$ as fluctuation coordinates as the enthalpy $H=H(S,P)$ is naturally a function of these coordinates in extended thermodynamic description of black holes~\cite{Xu:2019gqm,Ghosh:2019pwy}. We postpone a study of thermodynamic geometry of present case in these planes for future. The metric in  eqn. (\ref{metricG}) can be brought to the form:
\begin{equation}
 ds^2=\frac{1}{T} \left(-\frac{C_V}{T} dT^2+  \left(\frac{\partial P}{\partial V}\right)_{T} dV^2\right).
\end{equation}
Using the equation of state in eqn. (\ref{eos1}) for AdS massive gravity black hole system, it is possible to get an explicit form of the metric and compute the curvature. However, since we have the vanishing heat capacity at constant volume i.e., $C_V =0$, we employ the normalized scalar curvature $R_\text{N}$ defined in~\cite{Wei:2019uqg,Wei:2019yvs}, as:
\begin{equation}
R_\text{N} = \frac{(\partial_V P)^2 - T^2(\partial_{V, T} P)^2 + 2T^2(\partial_V P)(\partial_{V, T, T} P)}{2(\partial_V P)^2}\, ,
\end{equation}
which can be obtained easily in the present case.
Before analyzing this curvature further, we take a specific limit below to check consistency of our results and then discuss the general case.\\

{\underline {\bf Case-I: }} $c_1 =0$:
\vskip 0.5cm
From eqn. (\ref{RNc1}), the normalized scalar curvature in the special case $c_1=0$ is thus:
\begin{equation} \label{RN2}
R_\text{N} = \frac{\Big\{   9(k + m^2 c_2 c_0^2)^2 V^\frac{4}{3} - (k+ m^2 c_2 c_0^2)(36\pi V)^\frac{2}{3}(4q^2 + 3TV) + 8(6\pi^2)^\frac{2}{3}(2q^4 + 3q^2 TV)   \Big\}}{2\Big[3(k+m^2 c_2 c_0^2)V^{2/3} - (6\pi^2)^{1/3}(4q^2+3TV) \Big]^2}  \, .
\end{equation} 
\begin{figure}[h!]
	
	{\centering
		{\includegraphics[width=3.1in]{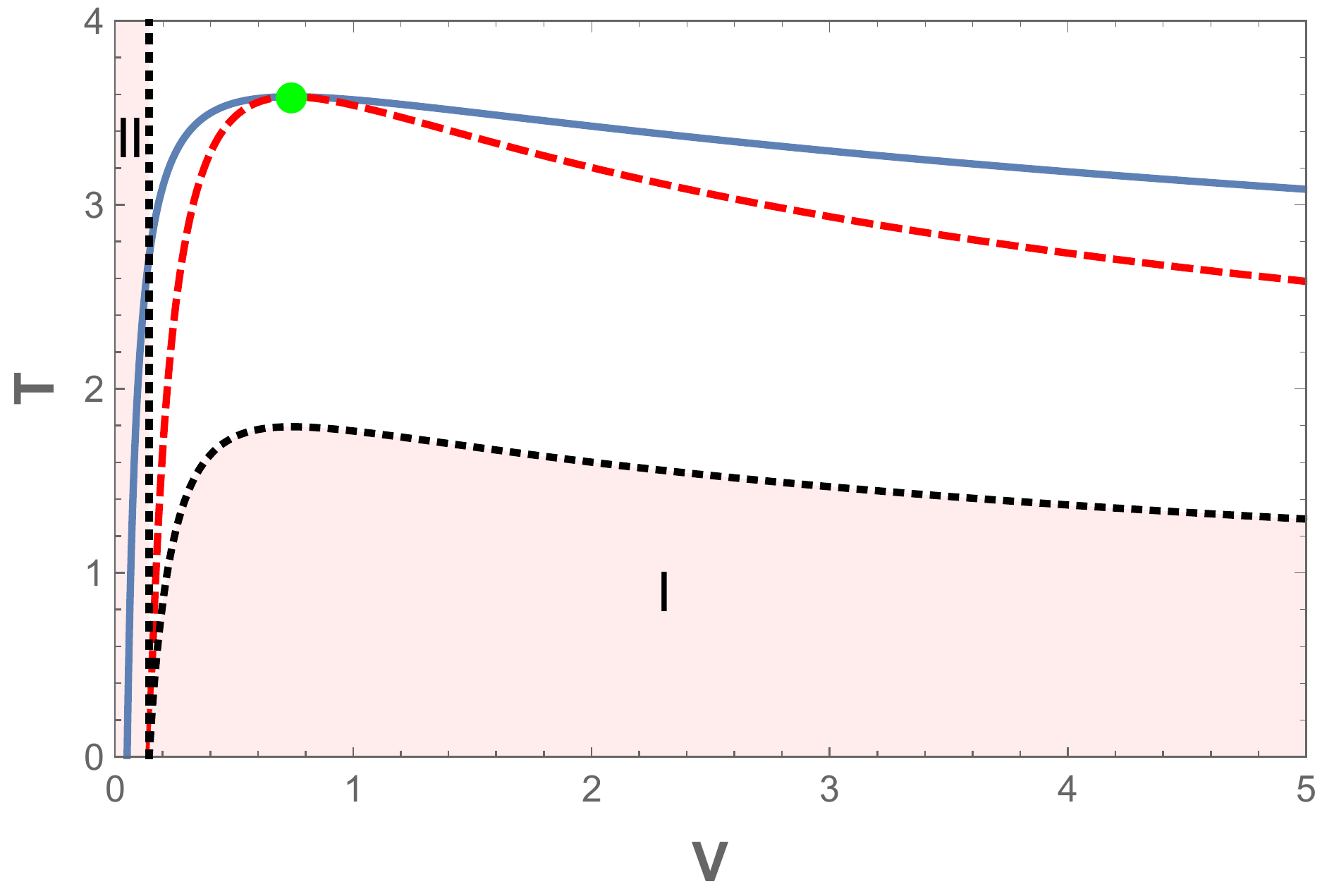}}

		\caption{\footnotesize The coexistence curve (blue color), spinodal curve (dashed red color), $R_\text{N}$ is positive in the shadow regions, zero on dashed black curves(sign-changing curves), and negative in the remaining region. Coexistence curve begins at  $V = 0.050578$, while the spinodal curve and sign-changing curves begin at $V =0.143055$.  Other fixed parameters take the values $c_1=0,k=-1,q=1,m=2,c_0=1,c_2=5$, but the main features are independent of these values.
		}   
		\label{fig:tv_region_plot}	}
\end{figure}

 \begin{figure}[h!]
	
	{\centering
		\subfloat[]{\includegraphics[width=2.4in]{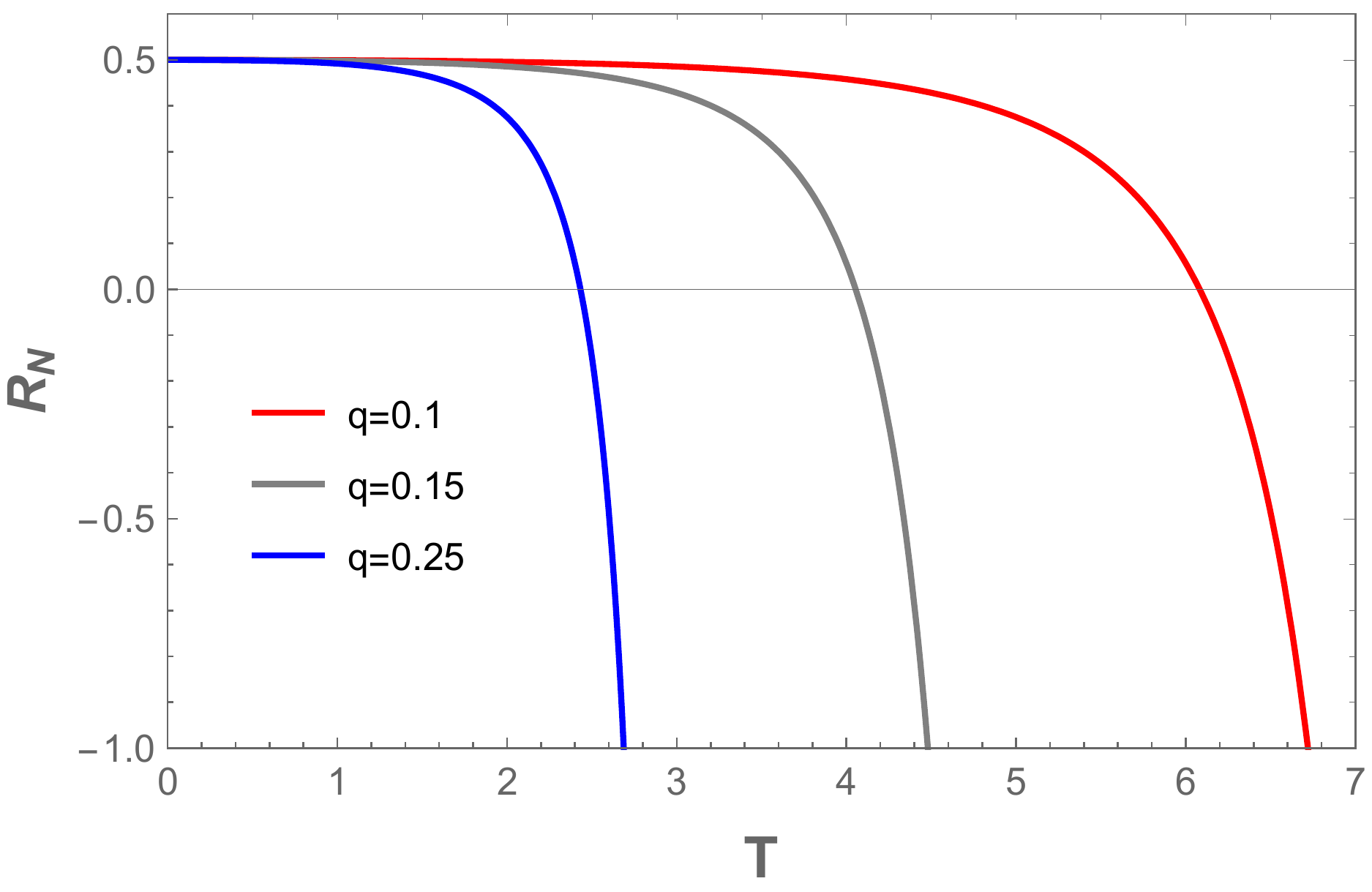}} \hspace{1.5cm}		
		\subfloat[]{\includegraphics[width=2.4in]{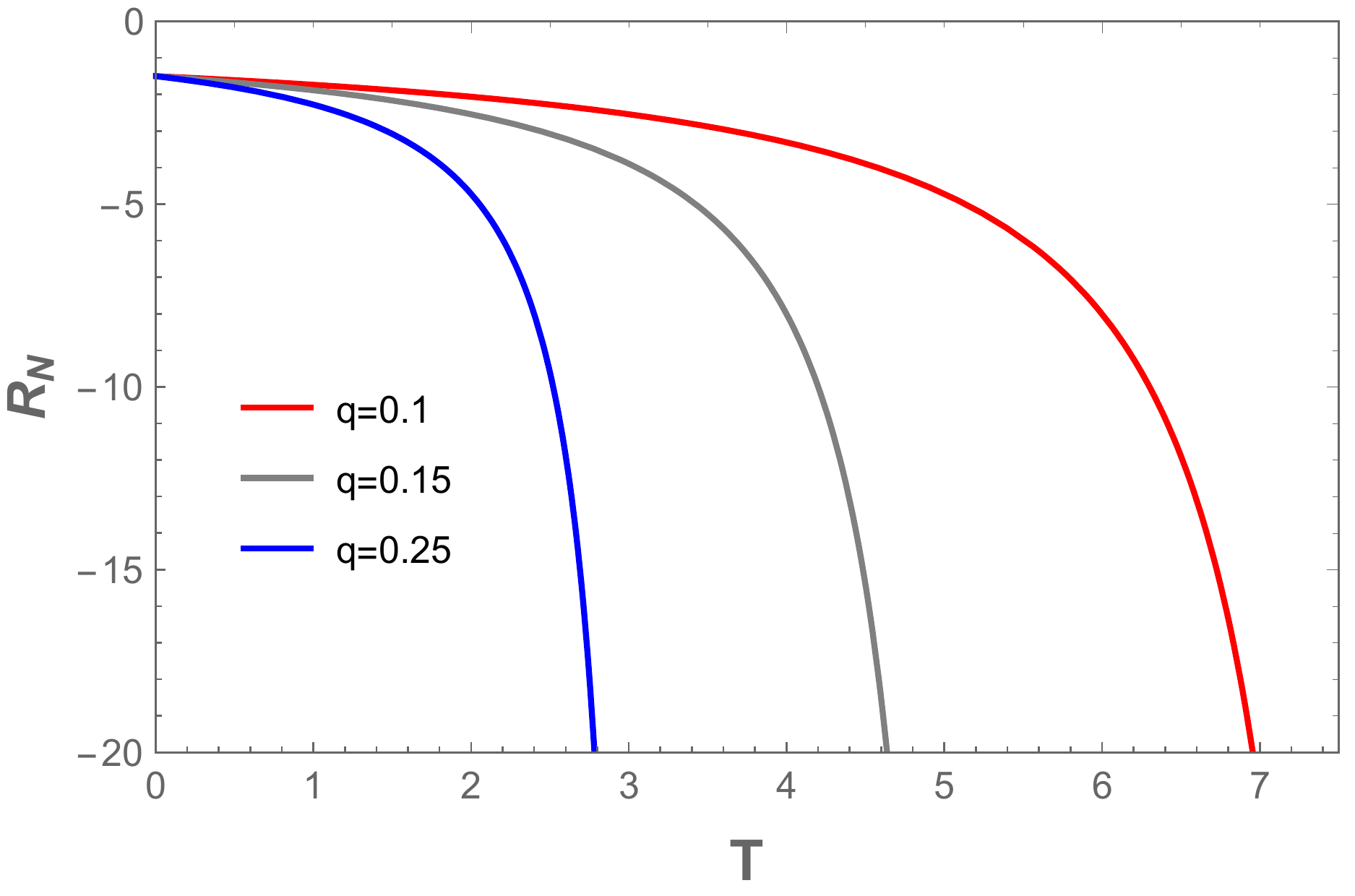}} \hspace{1.5cm}
		\subfloat[]{\includegraphics[width=2.4in]{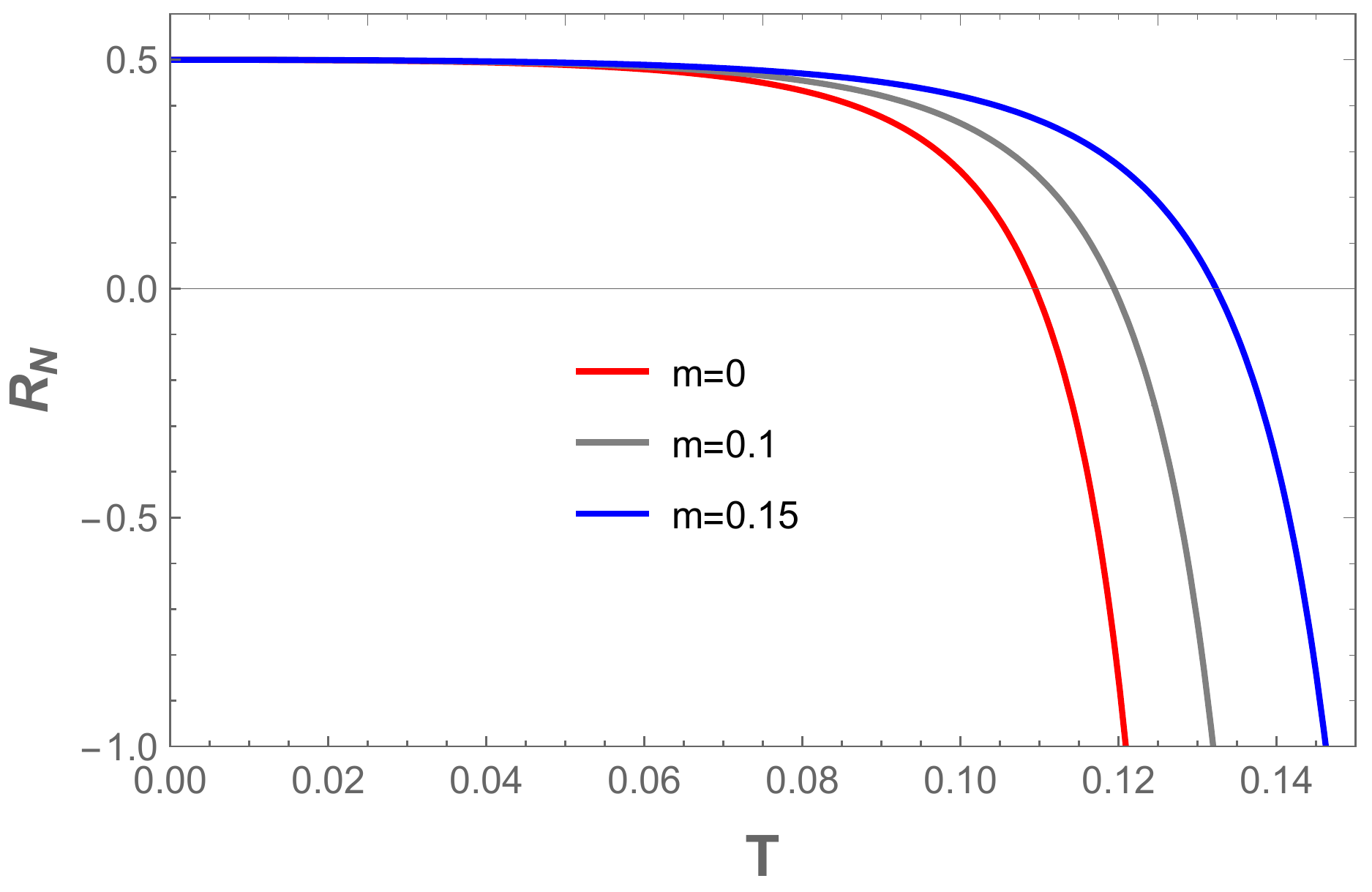}} \hspace{1.5cm}
		\subfloat[]{\includegraphics[width=2.4in]{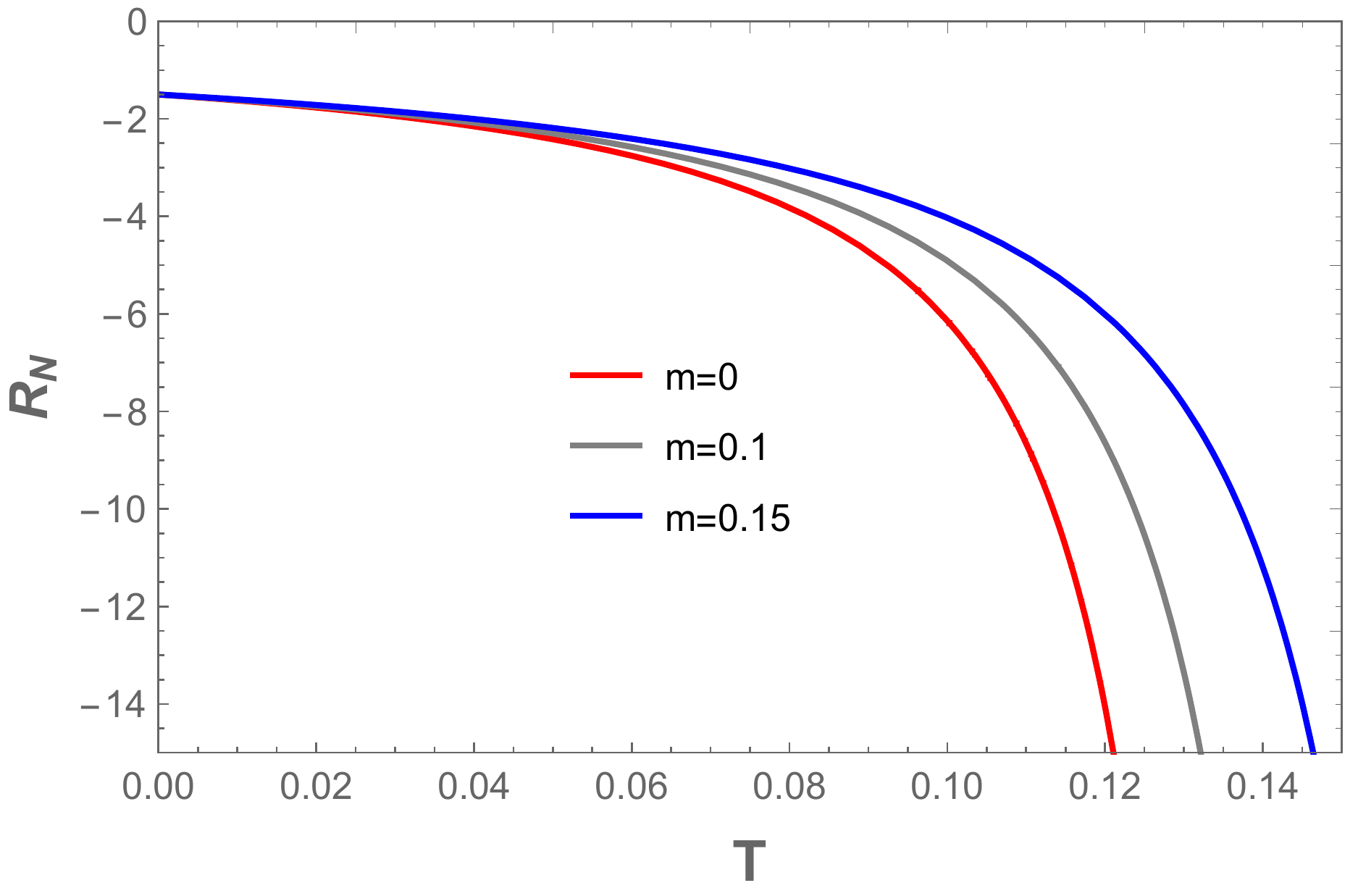}}\hspace{1.5cm}				
		\subfloat[]{\includegraphics[width=2.4in]{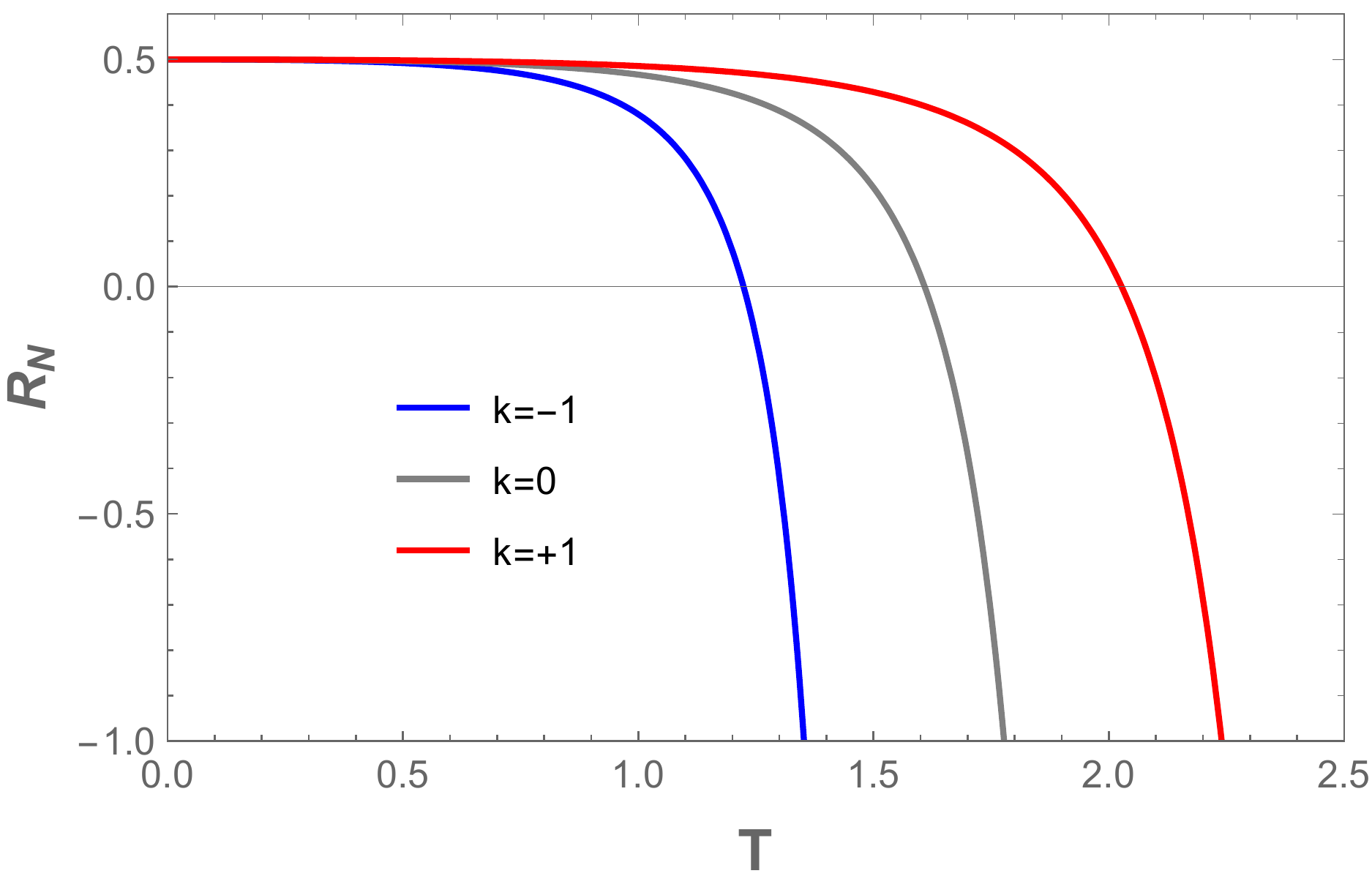}} \hspace{1.5cm}		
		\subfloat[]{\includegraphics[width=2.4in]{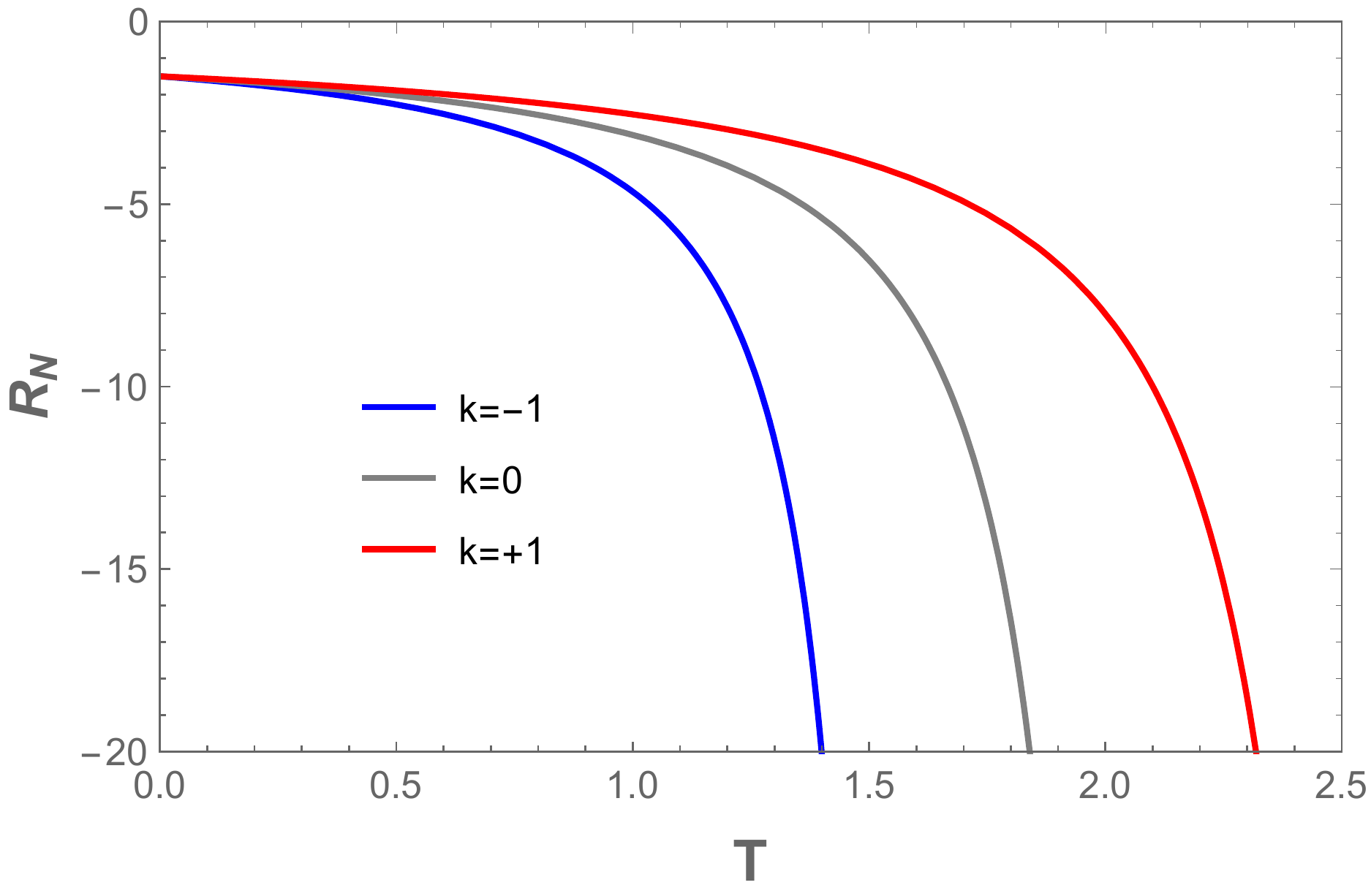}}

		\caption{\footnotesize Effects of various parameters on $R_\text{N}$ along the coexistence saturated small black hole curve (Left panel), and along the coexistence saturated large black hole curve (Right panel): Effect of charge $q$ in (a) and (b),  effect of graviton mass $m$ in (c) and (d), and effect of horizon topology $k$ in (e) and (f). (Here, we fixed the other parameters, $k=1, \  q=0.3, \  m=1, \  c_0 =2, \ c_2 = 1.5$. Similar behavior observed for other topologies as well).   
		}   
		\label{fig:non-reduced:1}	}
\end{figure}
As it turns out from the behavior of thermodynamic quantities for black hole in massive gravity at the critical point in~\eqref{red eq of st}, one can use the reduced equation of state (where by definition, dependence on all parameters of the theory, including black hole charge $q$ is scaled out) to proceed and map our system to the case in charged black holes in AdS~\cite{Wei:2019uqg,Wei:2019yvs} with identical results. However, since all the thermodynamic quantities (even at the critical point) intrinsically carry information about the massive gravity parameters, we continue in the non-reduced space, which is known to contain information about the effect of parameters of the model on the microstructures~\cite{Wei:2019ctz,Ghosh:2019pwy}.  First, the normalized scalar curvature $R_\text{N}$ diverges at the temperature:
\begin{equation}
T = T_\text{SP} =  \frac{ \epsilon(36\pi)^{1/3}V^{2/3}-8\pi q^2}{6\pi V}.
\end{equation}
i.e., along the spinodal curve and also diverges at the critical point $(V,T)=(V_c,T_c)$.  In addition, $R_\text{N}$ becomes zero  at $V=\frac{V_c}{3\sqrt{3}}$, and also at the temperature
\begin{equation}
T_0 = \frac{T_\text{sp}}{2}=\frac{ \epsilon(36\pi)^{1/3}V^{2/3}-8\pi q^2}{12\pi V}\, .
\end{equation}
Moreover, $R_\text{N}$ is positive for  $V < \frac{V_c}{3\sqrt{3}}$, while for  $V > \frac{V_c}{3\sqrt{3}}$, $R_\text{N}$ is positive when $T< T_0$ and negative when $T > T_0$.  In fact from Fig.~\eqref{fig:tv_region_plot}, one notes from the coexistence curve, spinodal curve, and the regions of  negative, zero and positive $R_\text{N}$ that: any behavior of $R_\text{N}$ (including divergent and positive) in the region below the coexistence curve must not be considered as the equation of state~\eqref{red eq of st} is invalid in this region. However, the positive region of $R_\text{N}$ above the coexistence curve(top part of region-II) indicates the domination of the repulsive interaction among the molecules of the black holes in massive gravity, a result first noted in the context of charged black holes in AdS in~\cite{Wei:2019uqg,Wei:2019yvs}.  Moreover, $R_\text{N}$ diverges at critical point for both the curves similar to the behavior of correlation length. 
At low temperature $\tilde{T}$, the coexistence saturated small black hole curve has positive $R_\text{N}$,  while the coexistence saturated large black hole curve has negative $R_\text{N}$, indicating that the repulsive interactions are dominating among the small black hole molecules and the attractive interactions are dominating among the large black hole molecules. 
This hints that  when a phase transition occurs at low $\tilde{T}$, the type of the interaction among the  molecules  also changes along with the size of the black hole, causing a huge change in the nature of microstructures.
Considering that the magnitude of $R_\text{N}$ is proportional to the intensity of the interaction, the large black hole molecules  are strongly interacting than  the small black hole molecules. Moreover, as the coexistence curve of small and large black holes is identical as that of the 4-dimensional charged AdS black hole in reduced parameter space~\cite{Wei:2019uqg,Wei:2019yvs}, one notes that the critical exponent of the $R_\text{N}$ near the critical point would be 2, and also the quantity $R_\text{N}(1-\tilde{T})^2$ (i.e., $R(1-\tilde{T})^2 C_v$) will take the universal value $-1/8$ in the limit $\tilde{T} \rightarrow 1$.\\

\begin{figure}[h!]
	
	{\centering
		
		\subfloat[]{\includegraphics[width=2.4in]{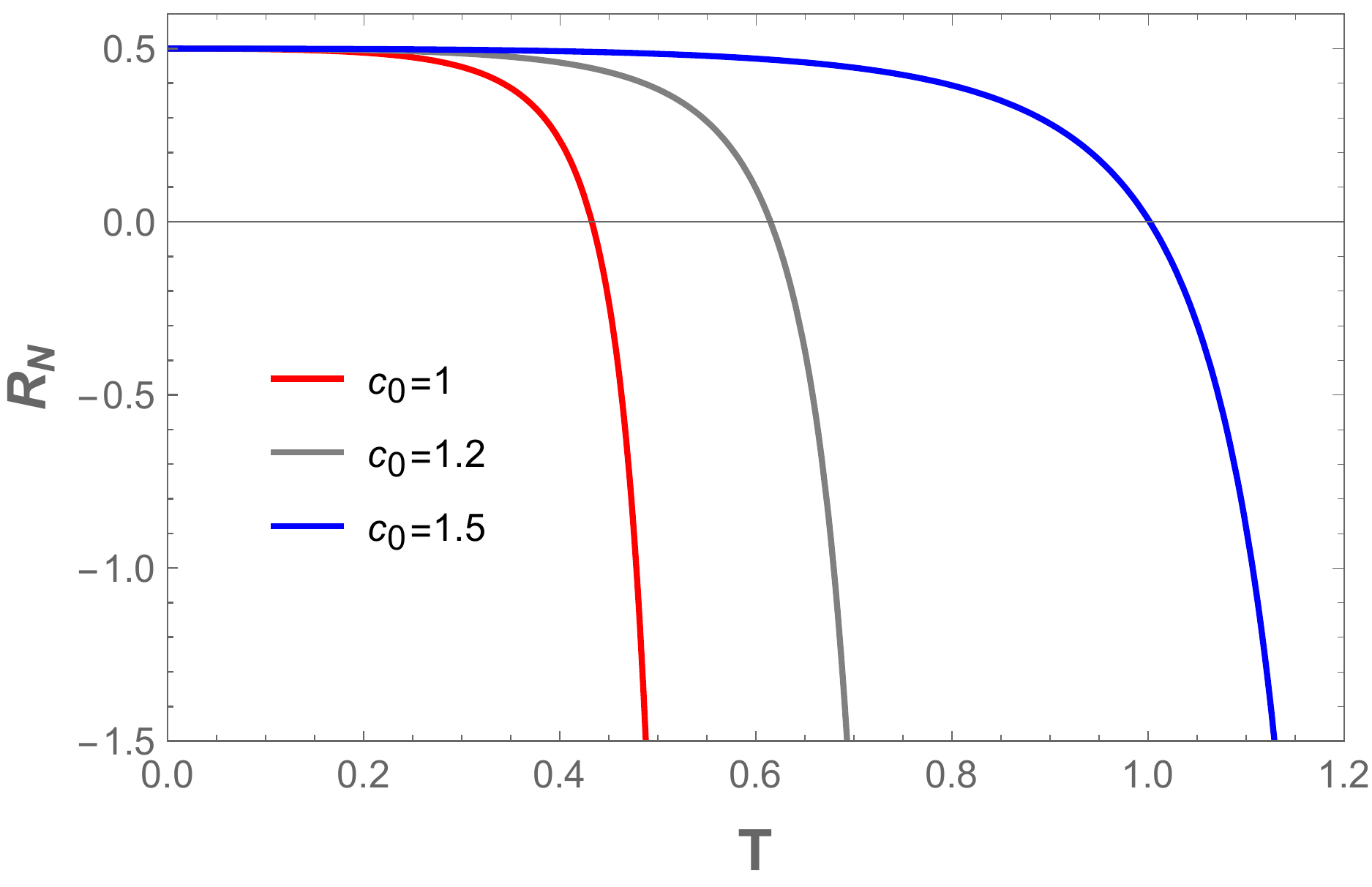}} \hspace{1.5cm}
		\subfloat[]{\includegraphics[width=2.4in]{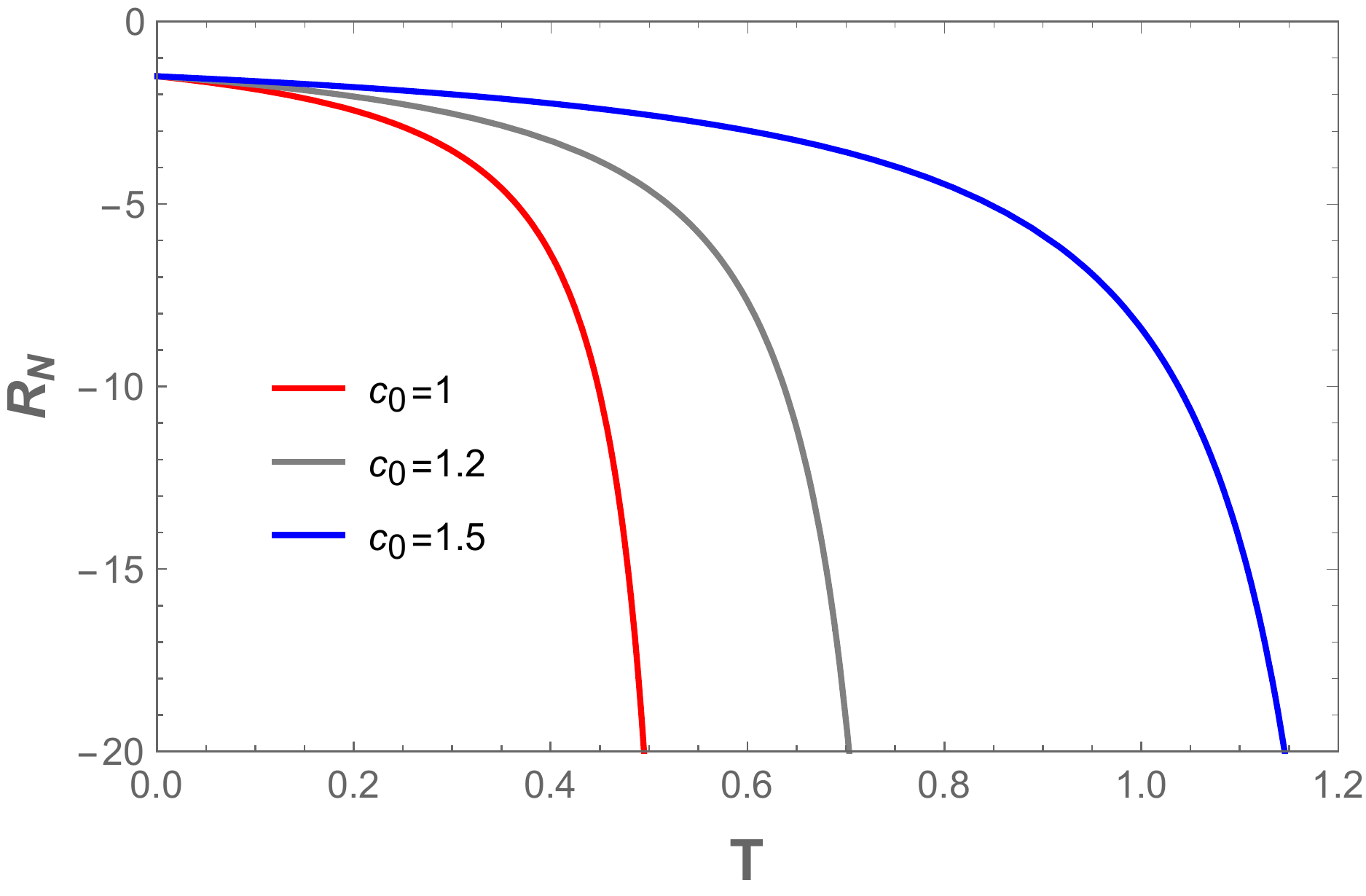}}	\hspace{1.5cm}			
		\subfloat[]{\includegraphics[width=2.4in]{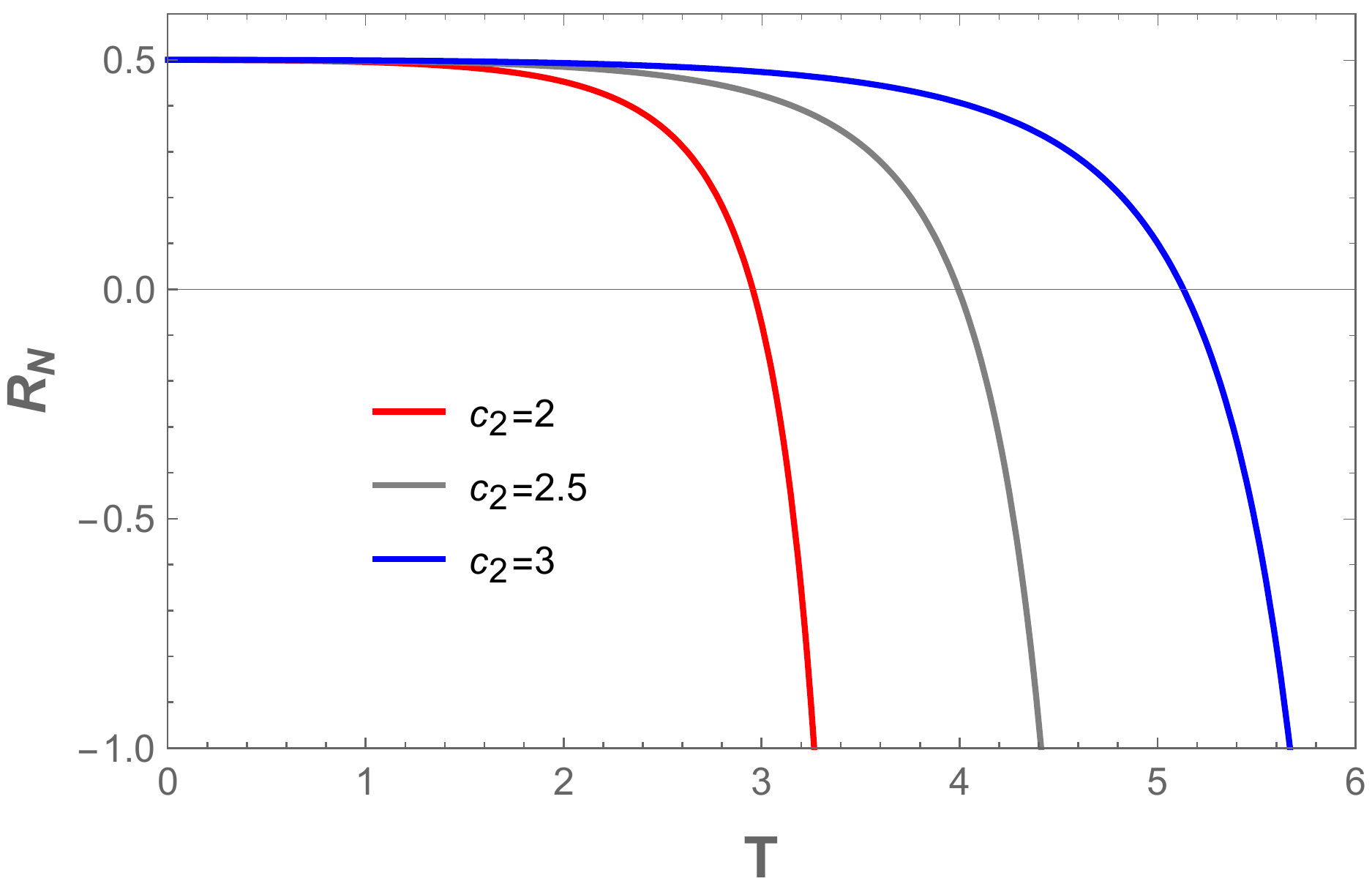}} \hspace{1.5cm}		
		\subfloat[]{\includegraphics[width=2.4in]{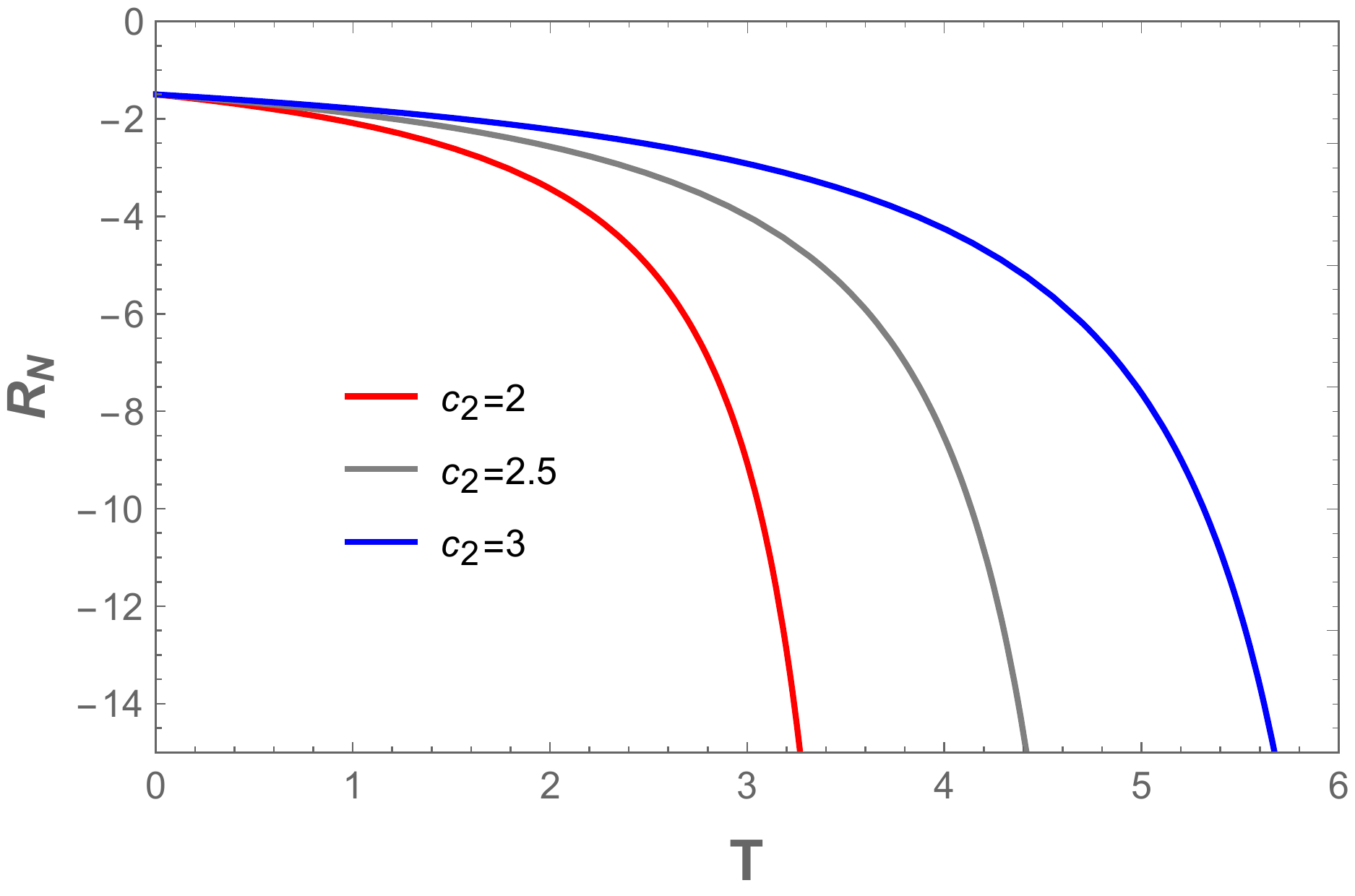}} \hspace{1.5cm}
		\subfloat[]{\includegraphics[width=2.4in]{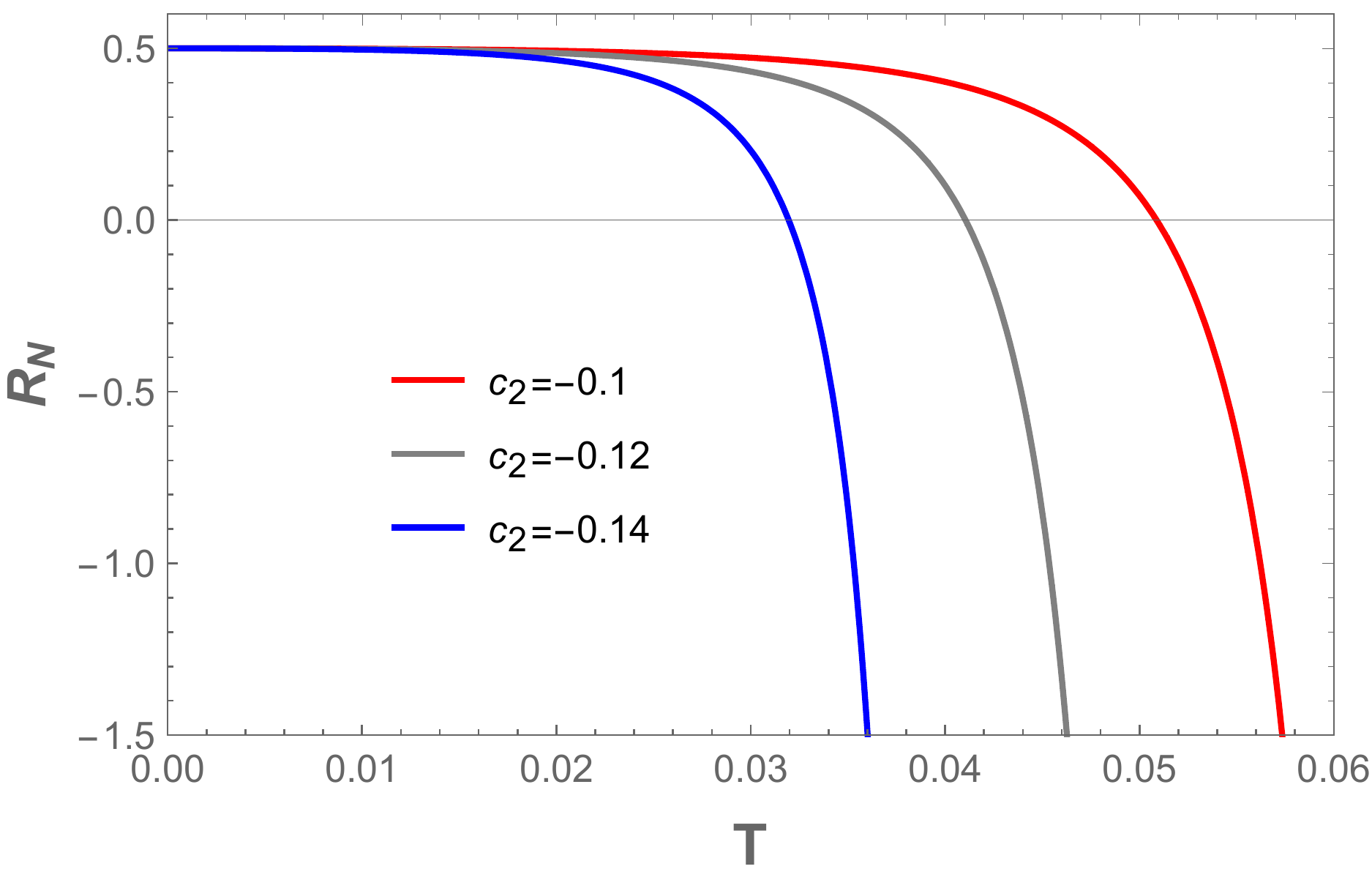}} \hspace{1.5cm}
		\subfloat[]{\includegraphics[width=2.4in]{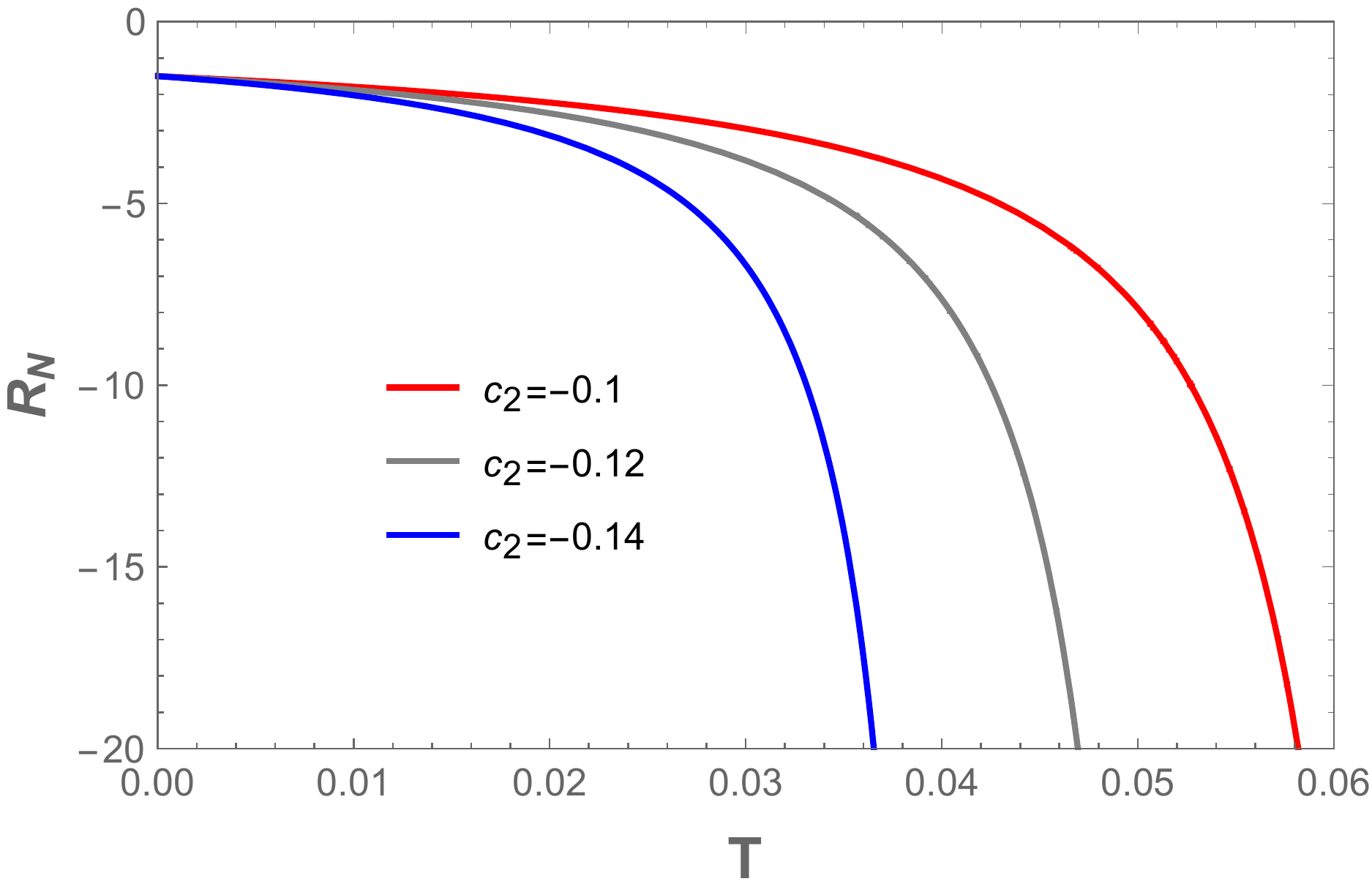}}

		\caption{\footnotesize Effects of various parameters on $R_\text{N}$ along the coexistence saturated small black hole curve (Left panel), and along the coexistence saturated large black hole curve (Right panel): Effect of massive coefficient $c_0$ in (a) and (b),  effect of  massive coefficient $c_2$ when positive valued in (c) and (d), and effect of  massive coefficient $c_2$ when negative valued in (e) and (f). (Here, we fixed the other parameters, $k=1, \  q=0.3, \  m=1, \  c_0 =2, \ c_2 = 1.5$. Similar behavior observed for other topologies as well).   
		}   
		\label{fig:non-reduced:2}	}
\end{figure}

\noindent
We now proceed to understand the effect of massive gravity parameters and horizon topology on the microstructures of charged black holes. The normalized scalar curvature $R_\text{N}$ along the coexistence curve of small and large black holes in non-reduced parameter space is explicitly plotted in Figures~\eqref{fig:non-reduced:1} and \eqref{fig:non-reduced:2}, in various possible situations. From figures-~\eqref{fig:non-reduced:1} (a),(b) and \eqref{fig:non-reduced:1} (c),(d), we note that, as the graviton mass $m$ increases and as  the charge $q$ decreases, the repulsive interactions of the molecules of small black hole are in respective cases, strongly repulsive. In all the above noted four figures, we checked that
 the attractive interactions of the molecules of large black hole are weakly attractive irrespective of horizon topology.  The effect of horizon topology when all other parameters of the model are held fixed are shown in figures-~\eqref{fig:non-reduced:1} (e),(f).  From figure-~\eqref{fig:non-reduced:1} (e) one notes that, repulsive interactions of small black hole microstructures are stronger in case of spherical topology of horizon ($k=1$), following by flat  ($k=0$) and weaker for hyperbolic topology  ($k=-1$). Where as,  figure-~\eqref{fig:non-reduced:1}(f) shows that the attractive interactions of large black hole microstructures are stronger for hyperbolic topology, followed by flat and weaker for spherical topology. Furthermore, from all the plots in figure-\eqref{fig:non-reduced:2} it is quite clear that, the presence of massive graviton in Reissner-Nordstrom AdS black hole, makes the repulsive interactions of the molecules of the small black hole  more strongly repulsive, while, making the attractive interactions of the molecules of the large black hole  weakly attractive. \\

\begin{figure}[h!]
	
	{\centering
		\subfloat[]{\includegraphics[width=2.8in]{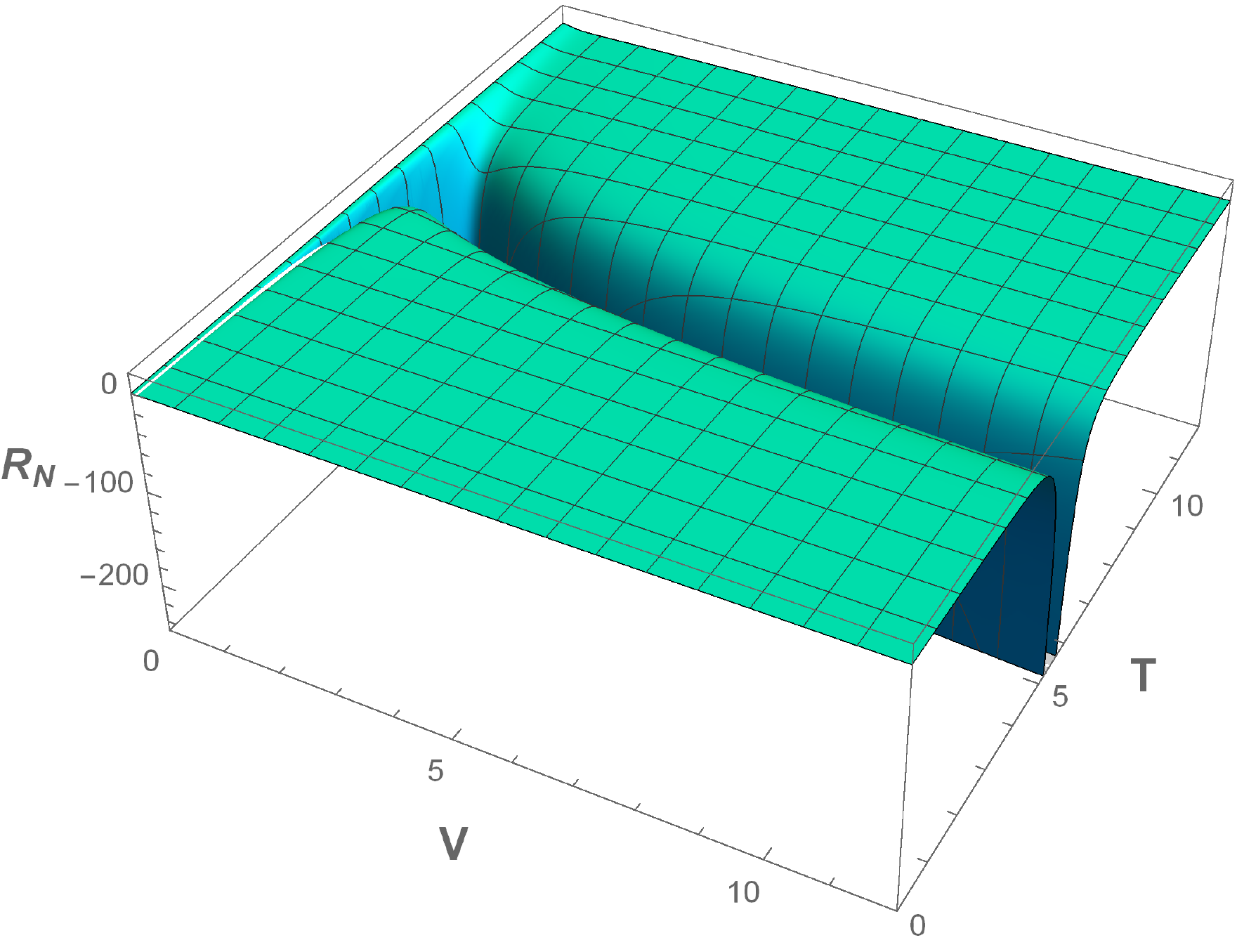}} \hspace{0.8cm}		
		\subfloat[]{\includegraphics[width=2.8in]{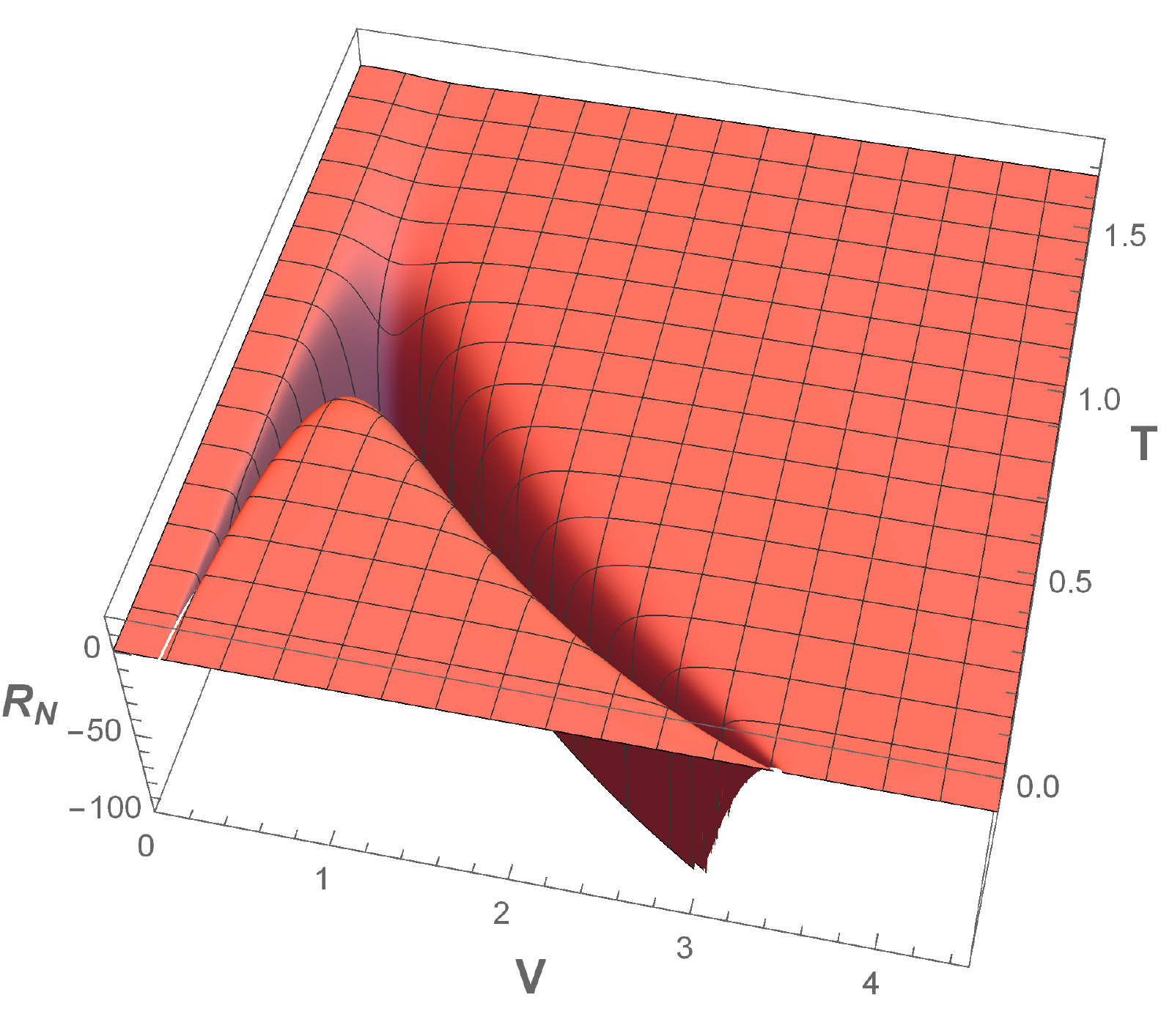}} \hspace{0.2cm}

		\caption{Spherical Topology case: \footnotesize  Behavior of the normalized scalar curvature $R_\text{N}$ with respect to $T$ and $V$ for the topological massive charged AdS black hole when the massive coefficient $c_1 \neq 0$. In (a) $c_1 = + 10$. (b) $c_1 = - 10$.  
	(Here, we set the other parameters $k=+1, \ q=1, \ m=2, \ c_0=1, \ c_2 =5.$)  	}   
		\label{fig:3D_k1_plots}	}
\end{figure}

\begin{figure}[h!]
	
	{\centering		
		\subfloat[]{\includegraphics[width=2.8in]{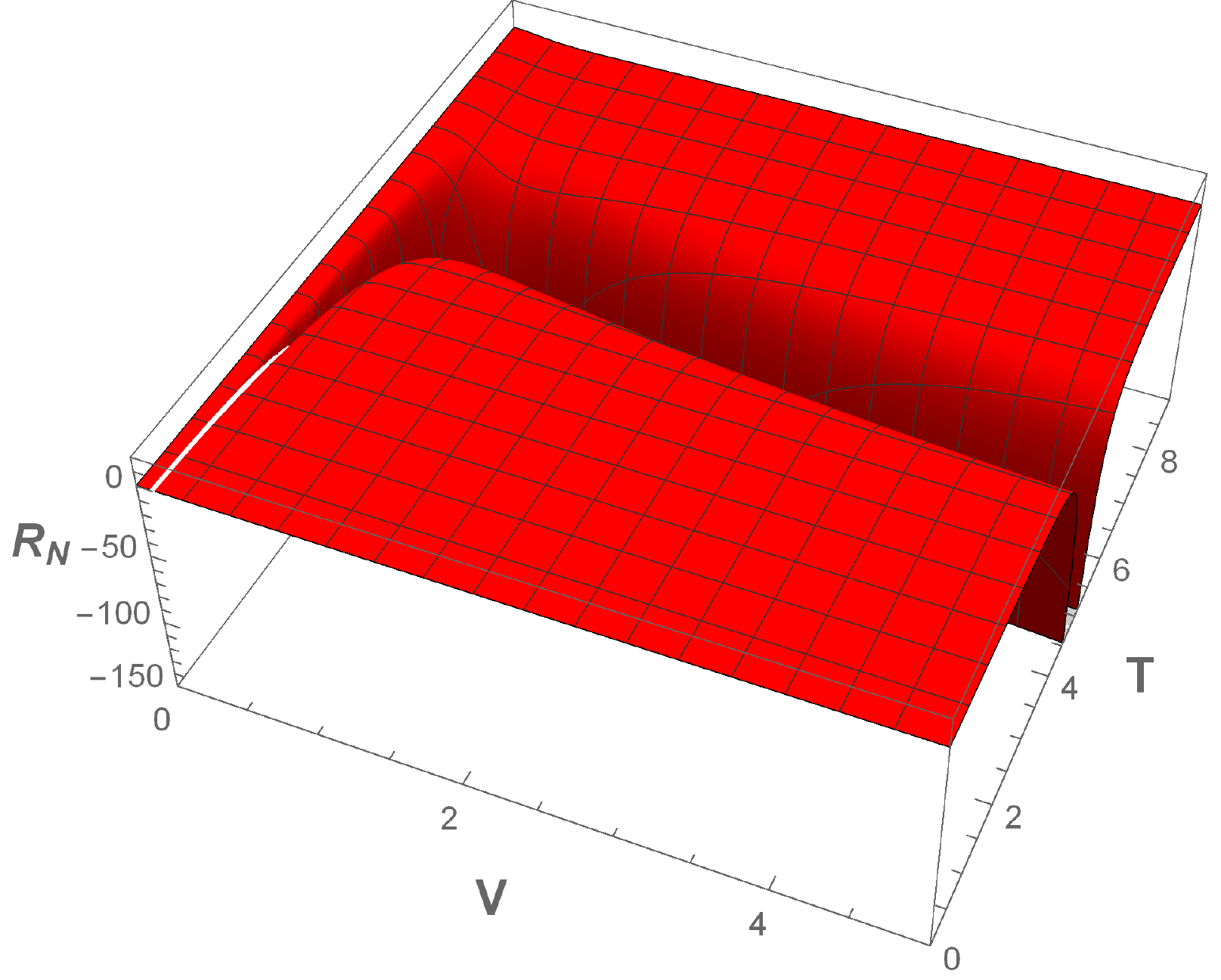}} \hspace{0.8cm}
		\subfloat[]{\includegraphics[width=2.8in]{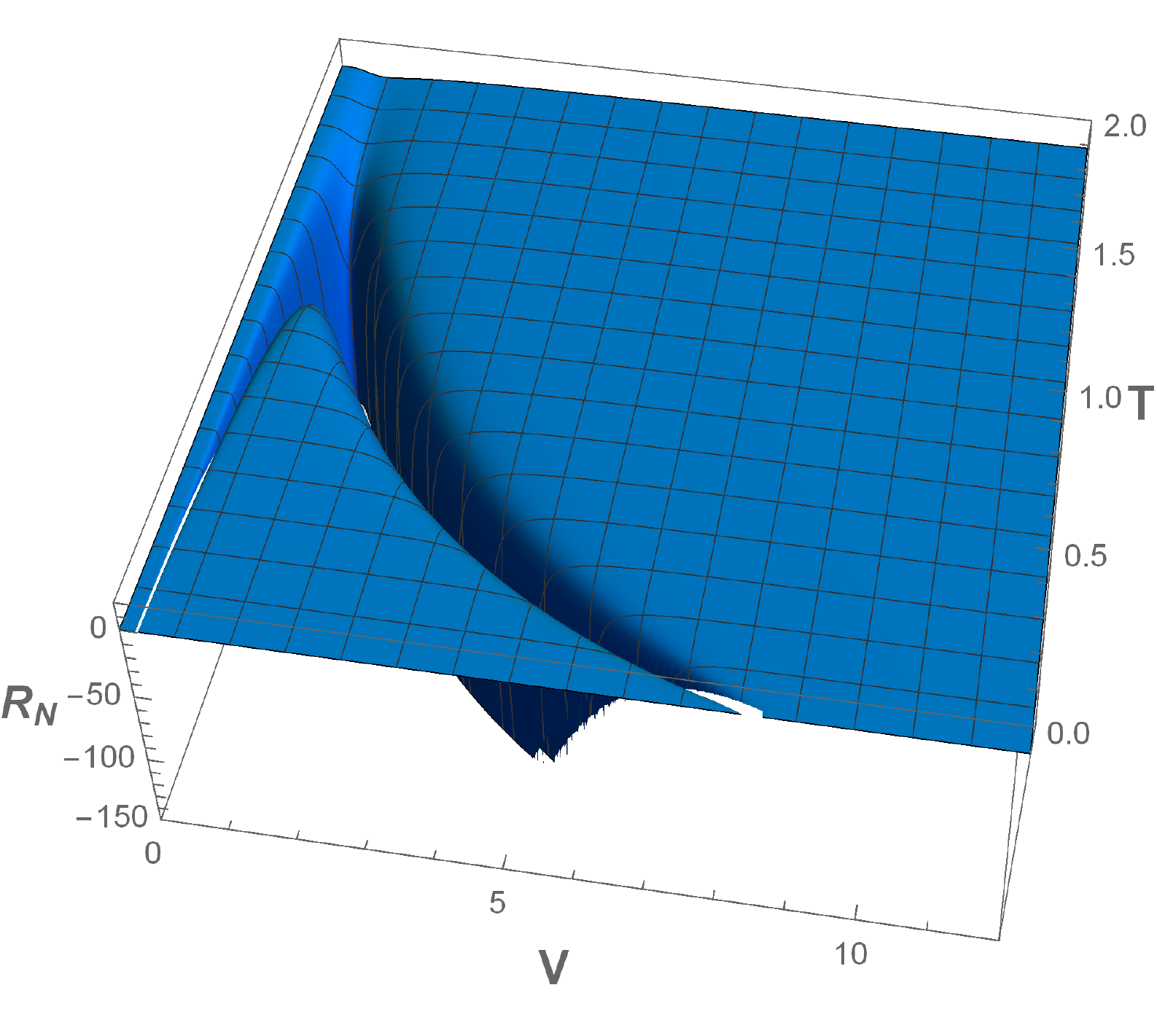}}		
				
		\caption{Hyperbolic Topology case: \footnotesize  Behavior of the normalized scalar curvature $R_\text{N}$ with respect to $T$ and $V$ for the topological massive charged AdS black hole when the massive coefficient $c_1 \neq 0$. In (a) $c_1 = +7$. (b) $c_1 = -7$.  
			(Here, we set the other parameters $ k=-1, \ q=1, \ m=2, \ c_0=1, \ c_2 =5.$)  	}   
		\label{fig:3D_km1_plots}	}
\end{figure}
\begin{figure}[h!]
	
	{\centering
		\subfloat[]{\includegraphics[width=2.75in]{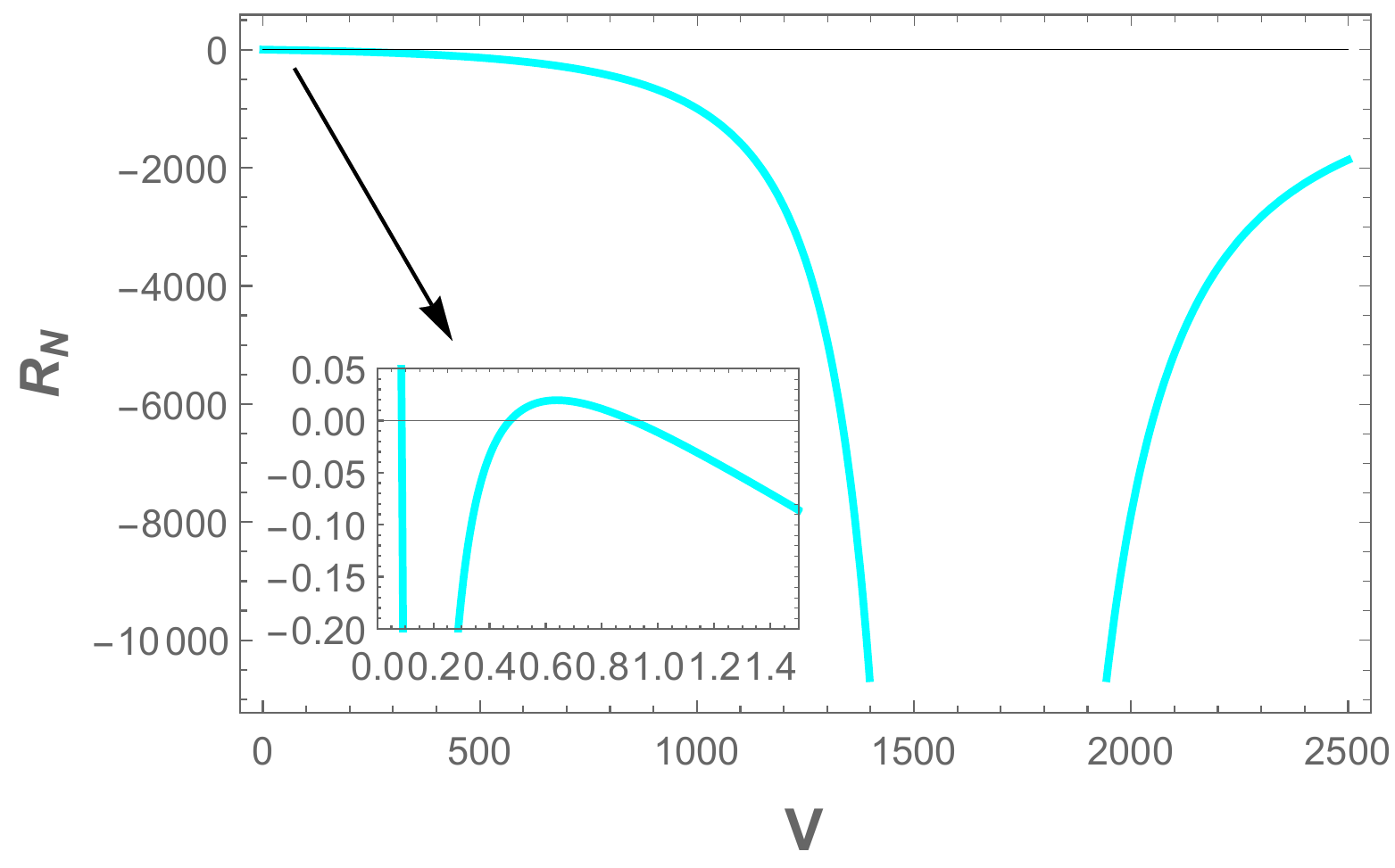}} \hspace{0.8cm}		
		\subfloat[]{\includegraphics[width=2.75in]{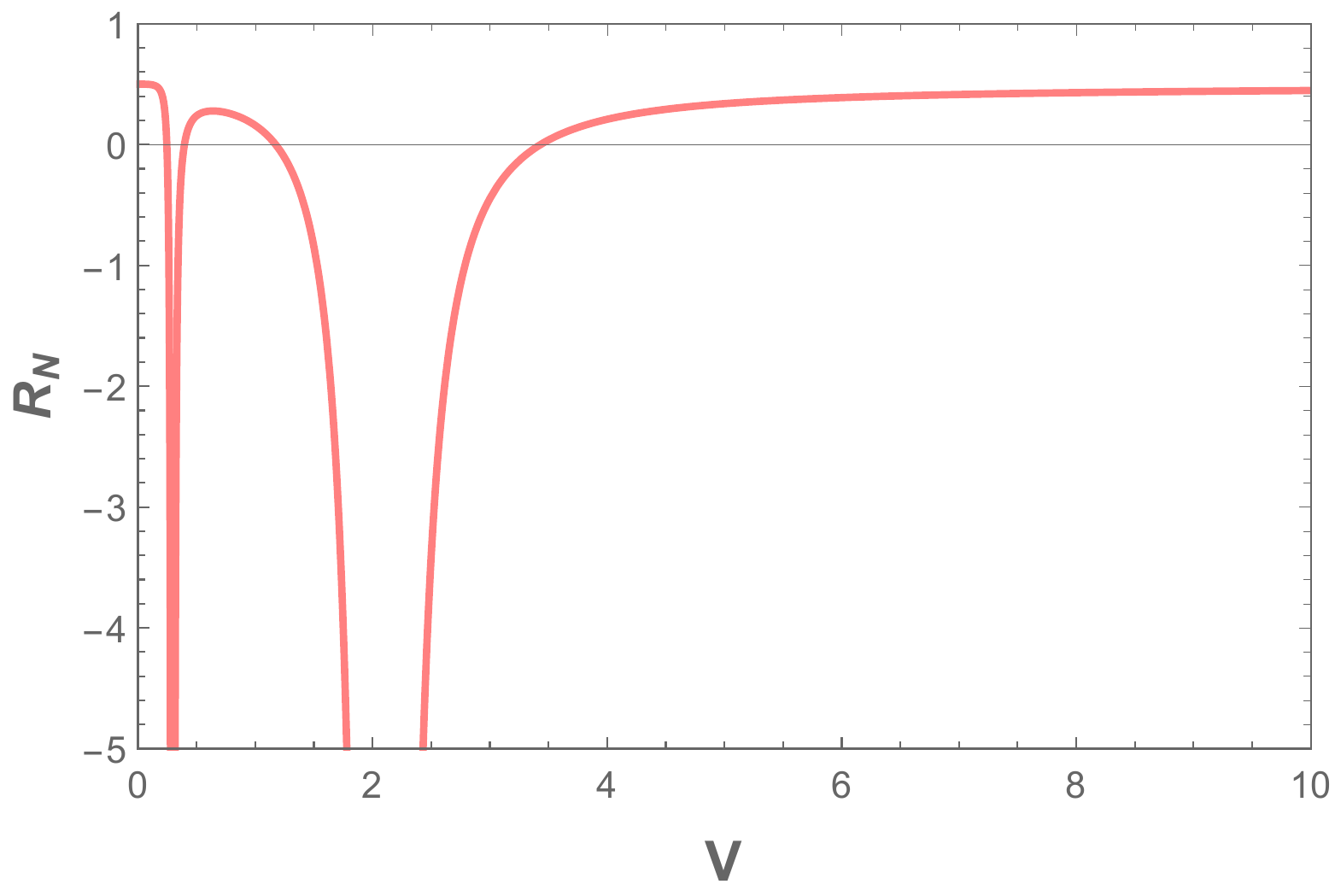}} \hspace{0.8cm}
		\subfloat[]{\includegraphics[width=2.75in]{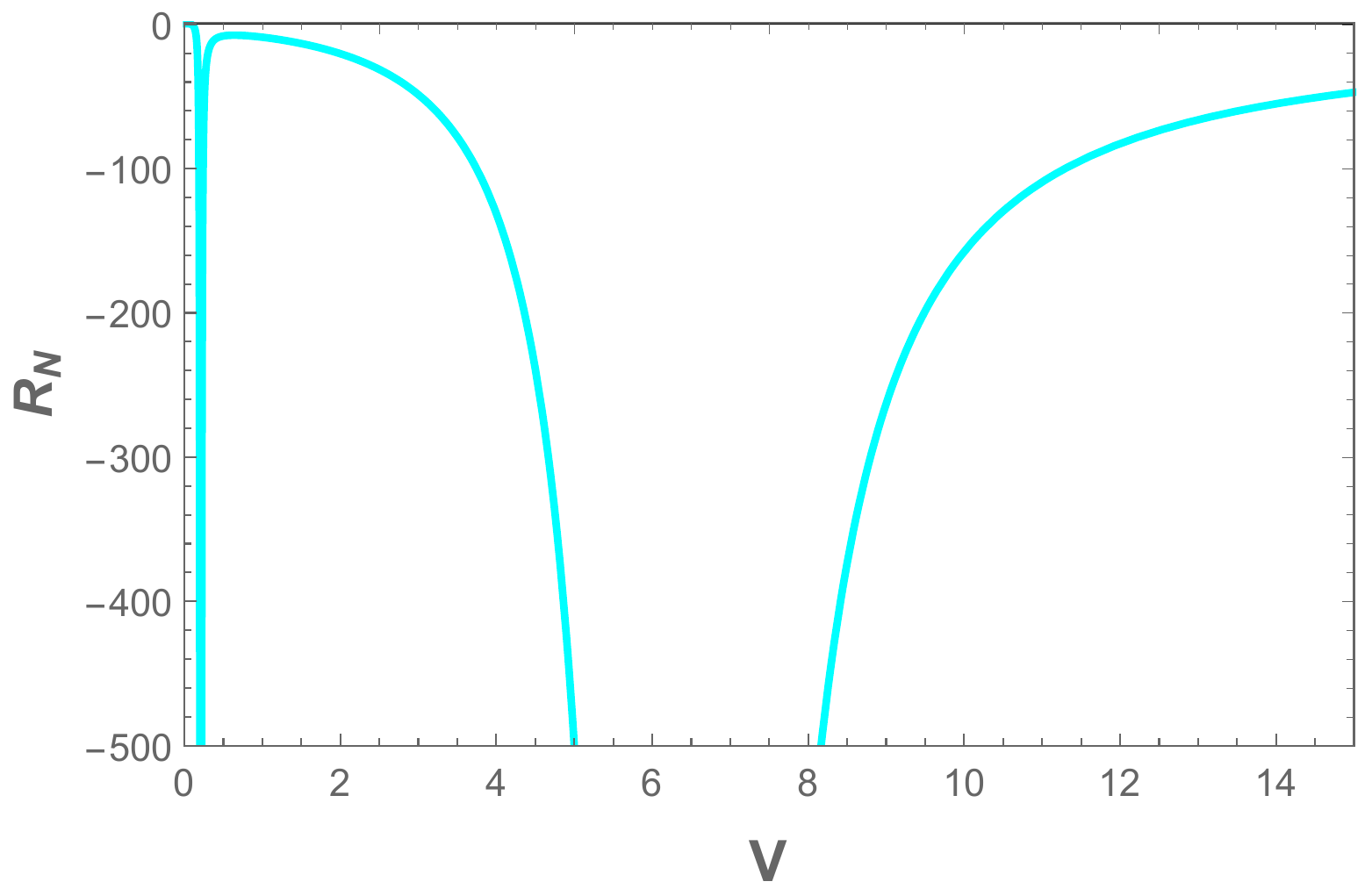}} \hspace{0.8cm}
		\subfloat[]{\includegraphics[width=2.75in]{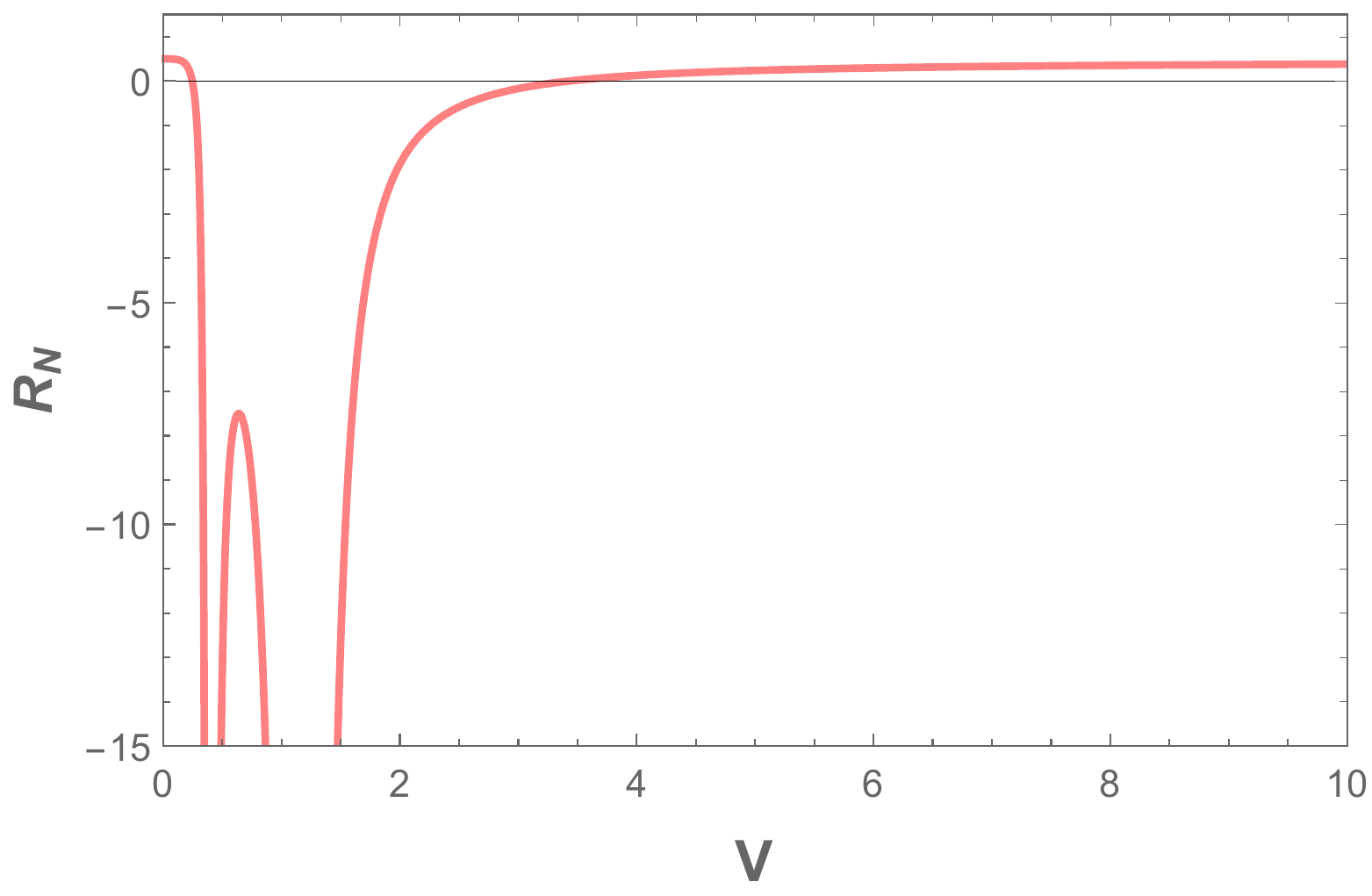}}\hspace{0.8cm}				
		\subfloat[]{\includegraphics[width=2.75in]{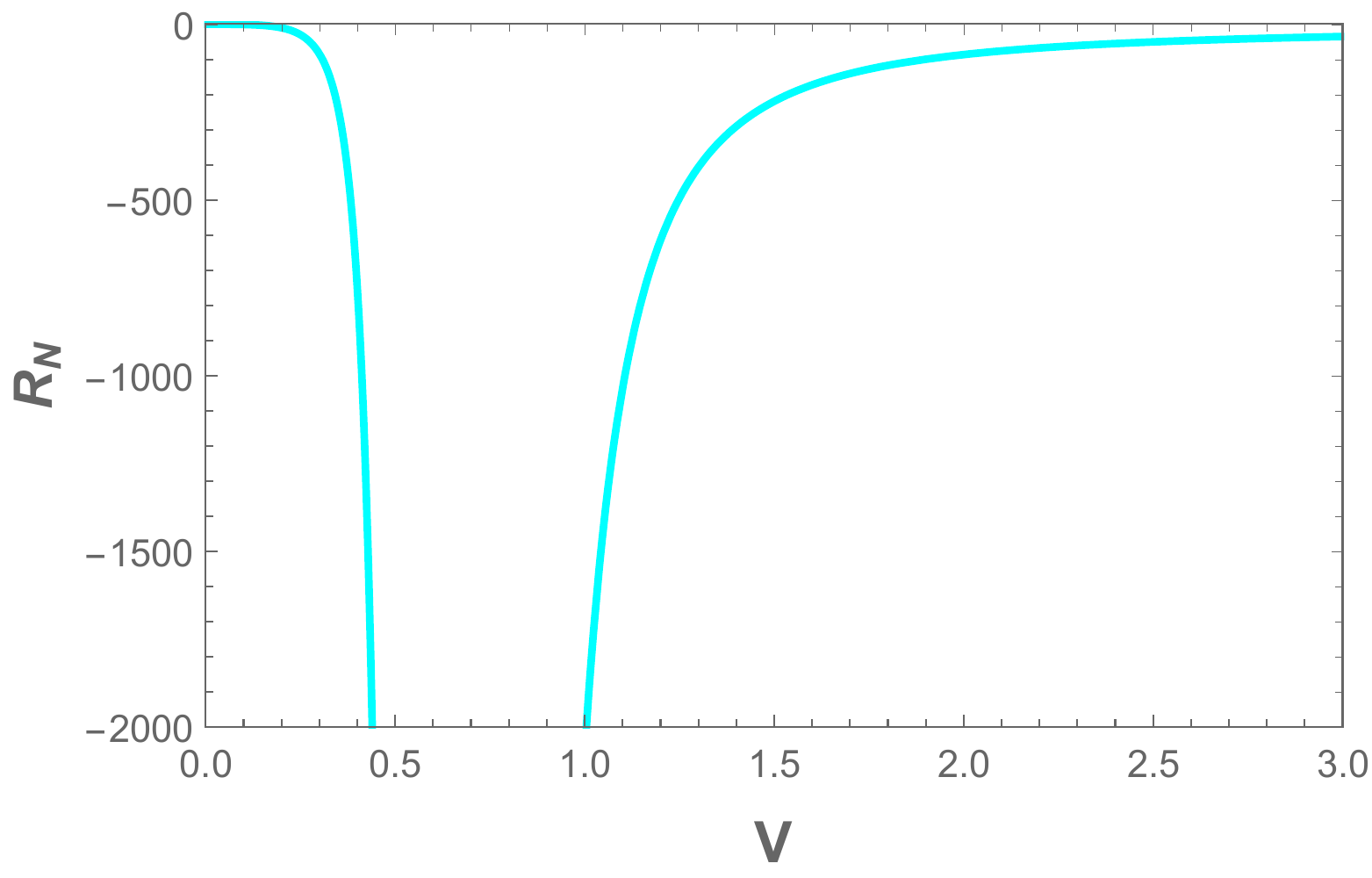}} \hspace{0.8cm}		
		\subfloat[]{\includegraphics[width=2.75in]{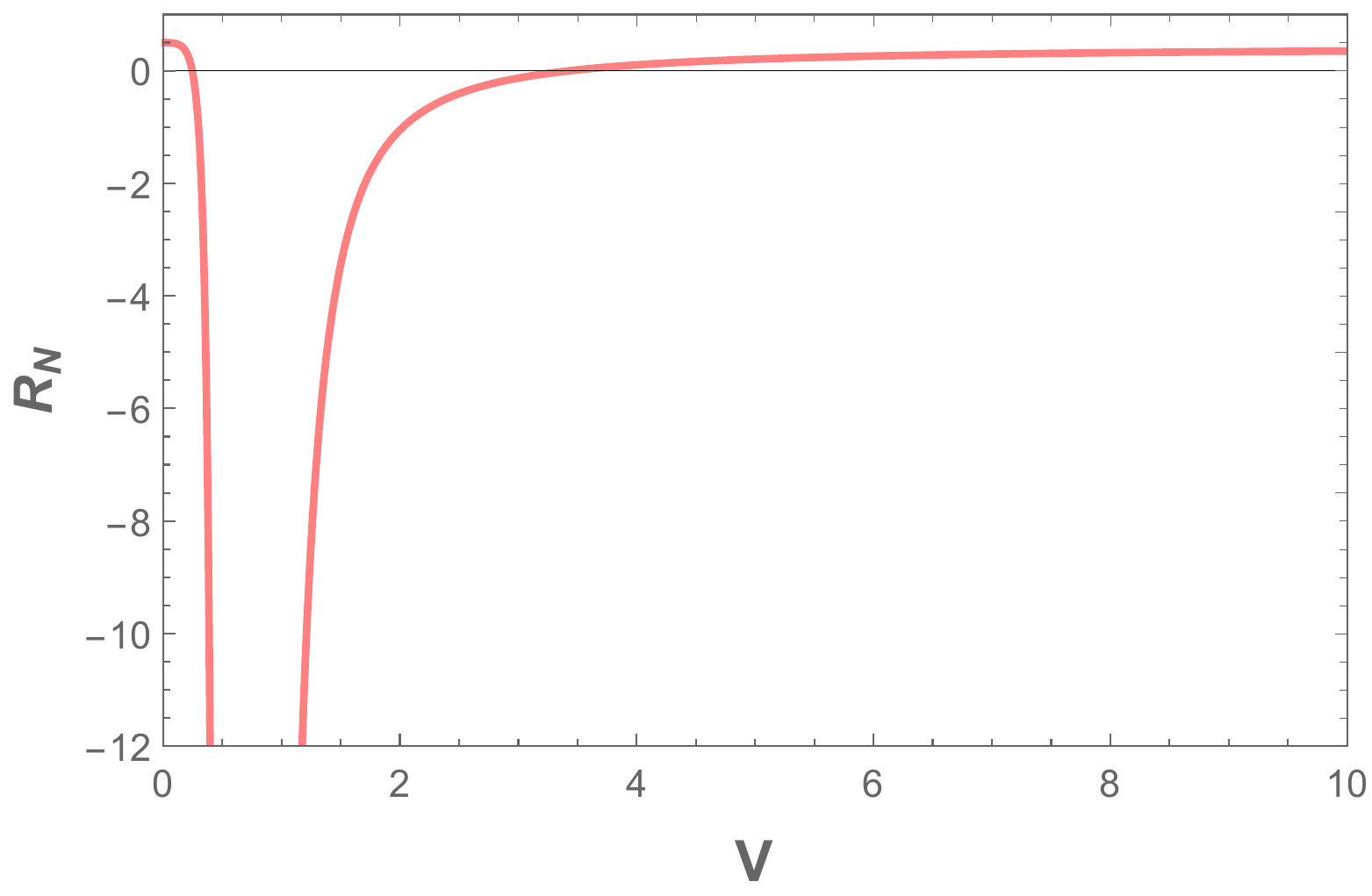}} \hspace{0.8cm}
		\subfloat[]{\includegraphics[width=2.75in]{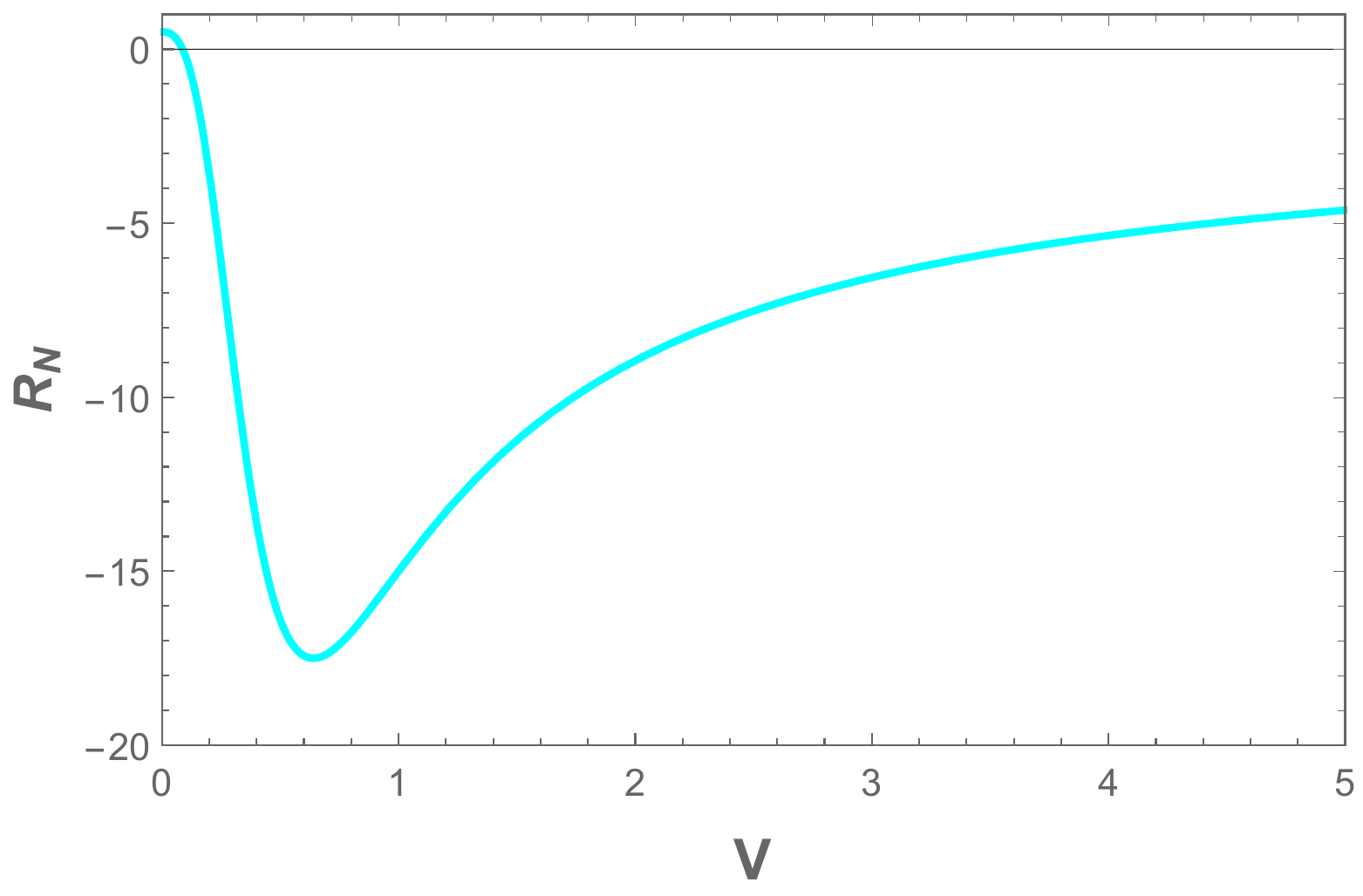}} \hspace{0.8cm}
		\subfloat[]{\includegraphics[width=2.75in]{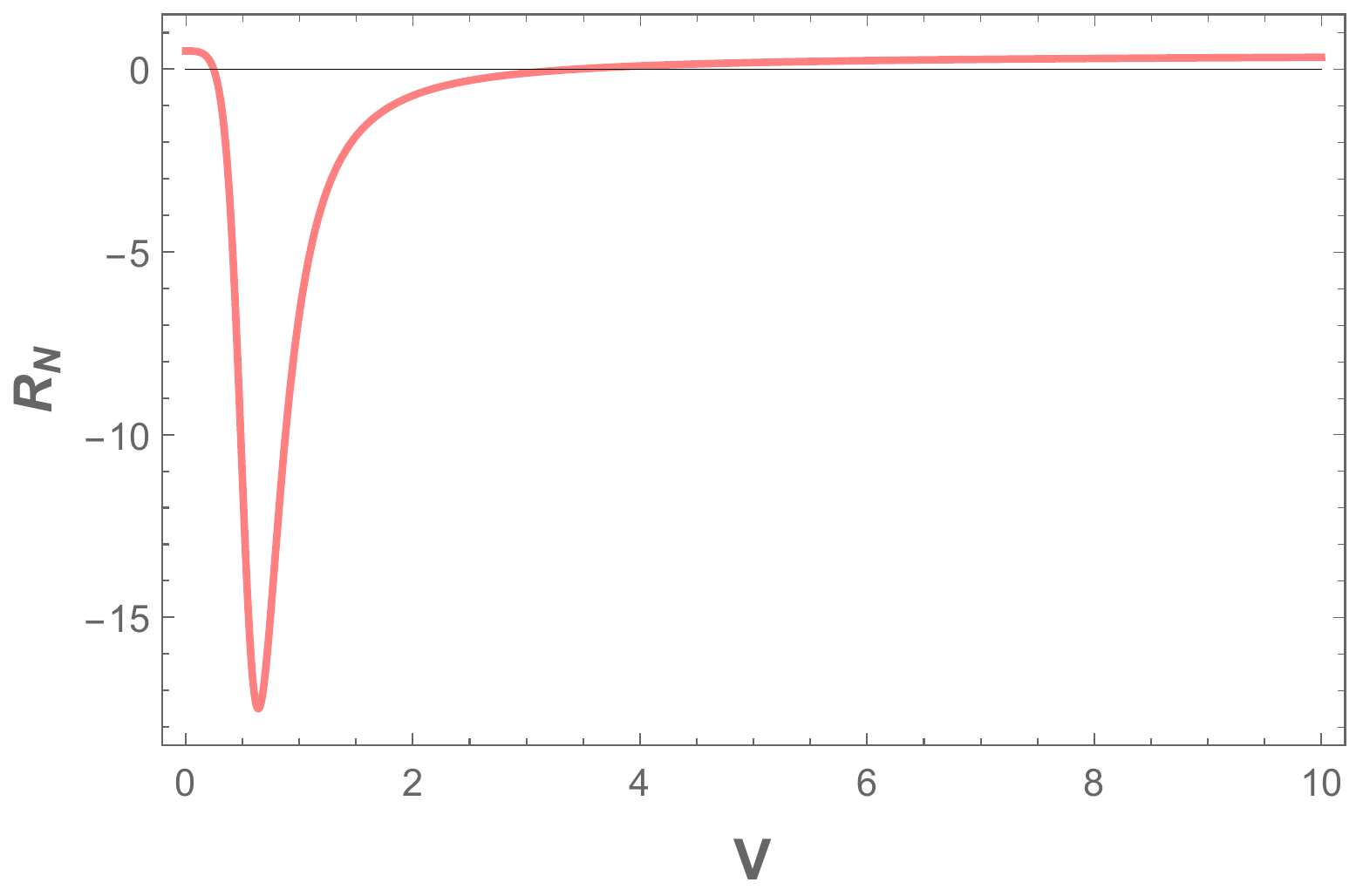}}

		\caption{Spherical Topology case: \footnotesize   Behavior of $R_\text{N}$ with respect to $V$ at fixed $T$ : $c_1 = +10$ in left panel(cyan colored) and $c_1 = -10$ in right panel(pink colored).   (a)  $T = 0.495T_c$ (Inset shows the positive value of $R_\text{N}$ in some region of $V$). (b)  $T = 0.4T_c$.   $T =0.8T_c$ in (c) and (d). $T = T_c$ in (e) and (f).   $T =1.2T_c$ in (g) and (h).  
			(Here, we set the other parameters $ k=+1, \ q=1, \ m=2, \ c_0=1, \ c_2 =5.$)	}   
		\label{fig:RN_fixed_T_k1_plots}	}
\end{figure}
\begin{figure}[h!]
	
	{\centering
		\subfloat[]{\includegraphics[width=2.75in]{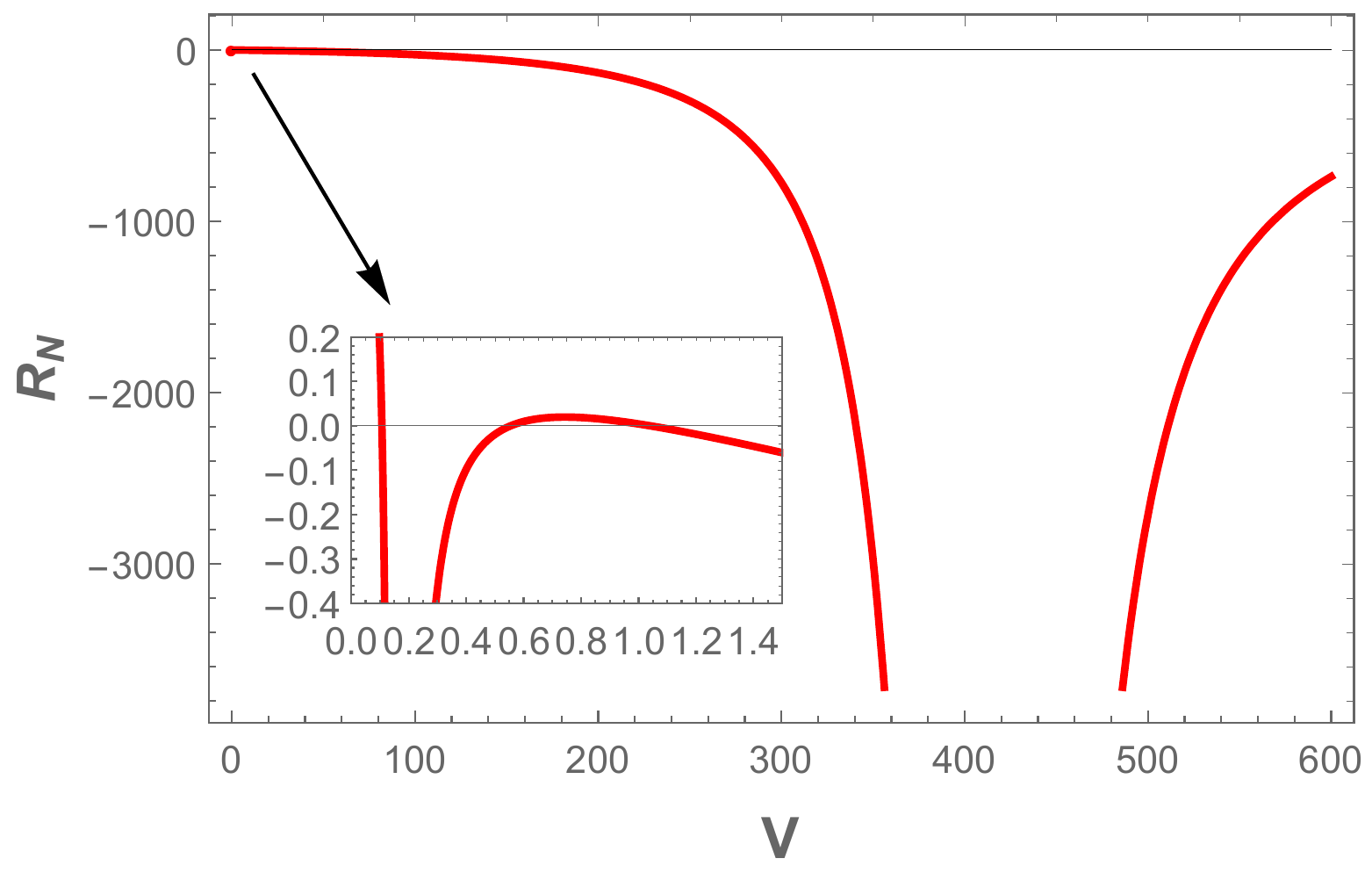}} \hspace{0.8cm}		
		\subfloat[]{\includegraphics[width=2.75in]{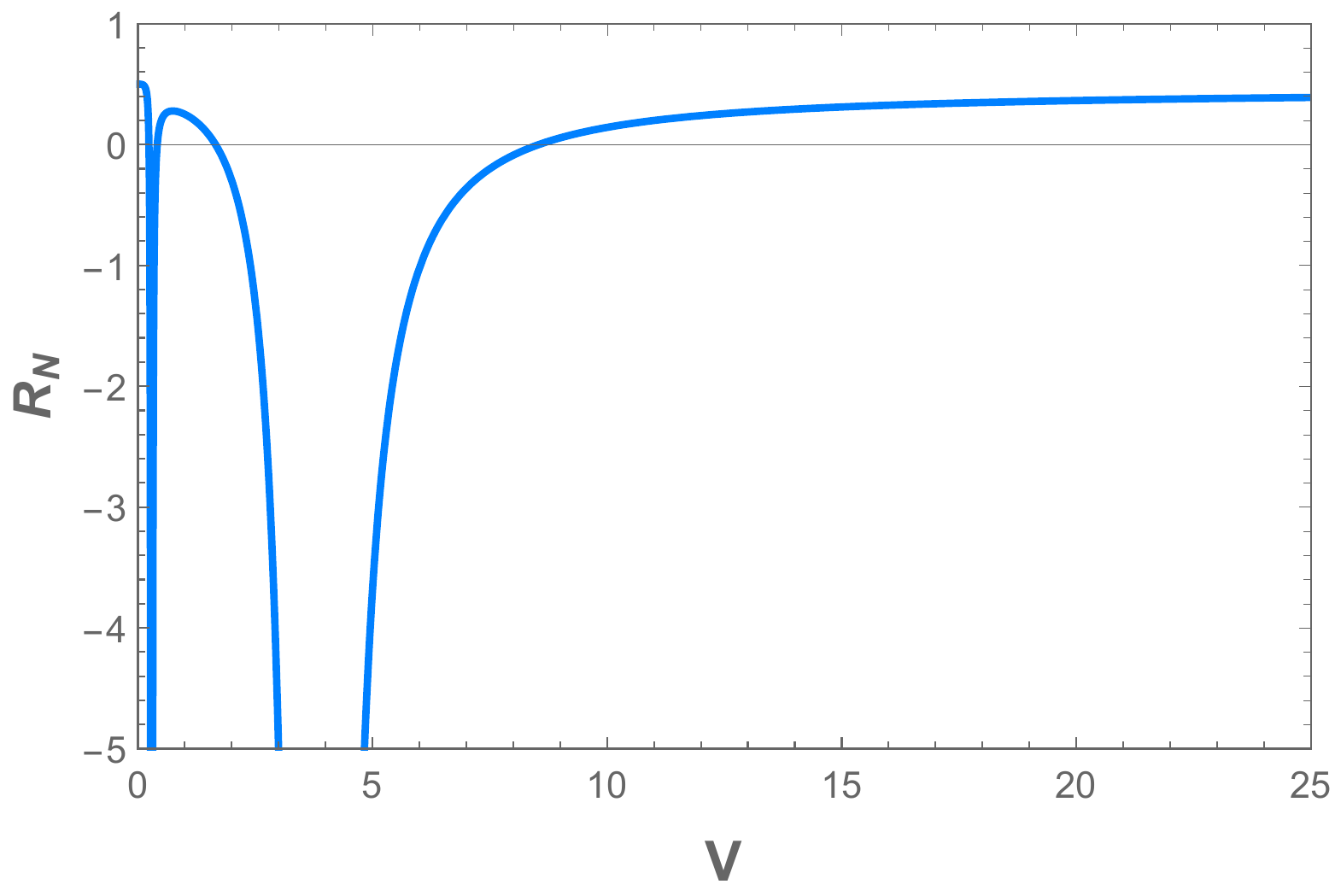}} \hspace{0.8cm}
		\subfloat[]{\includegraphics[width=2.75in]{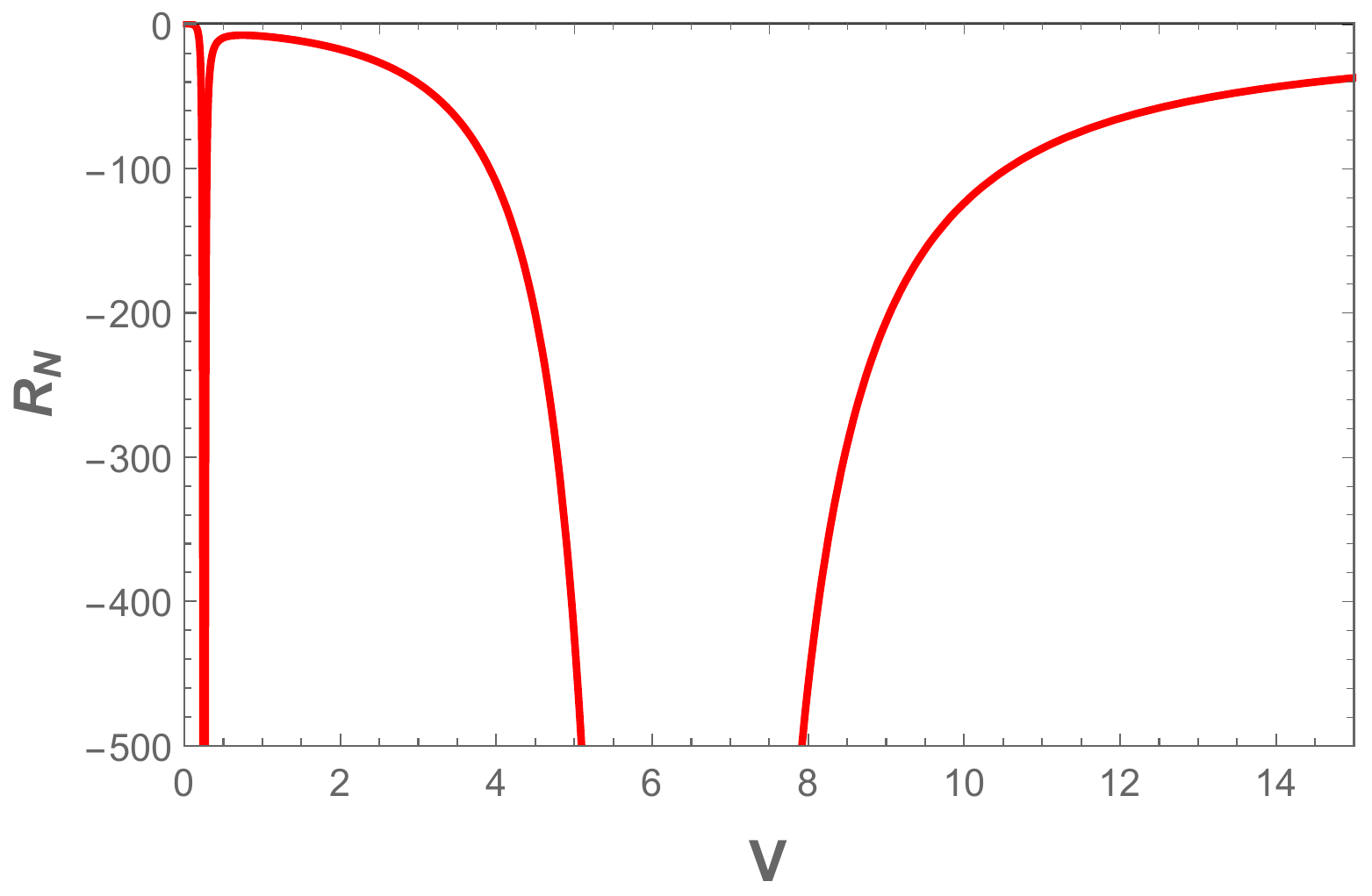}} \hspace{0.8cm}
		\subfloat[]{\includegraphics[width=2.75in]{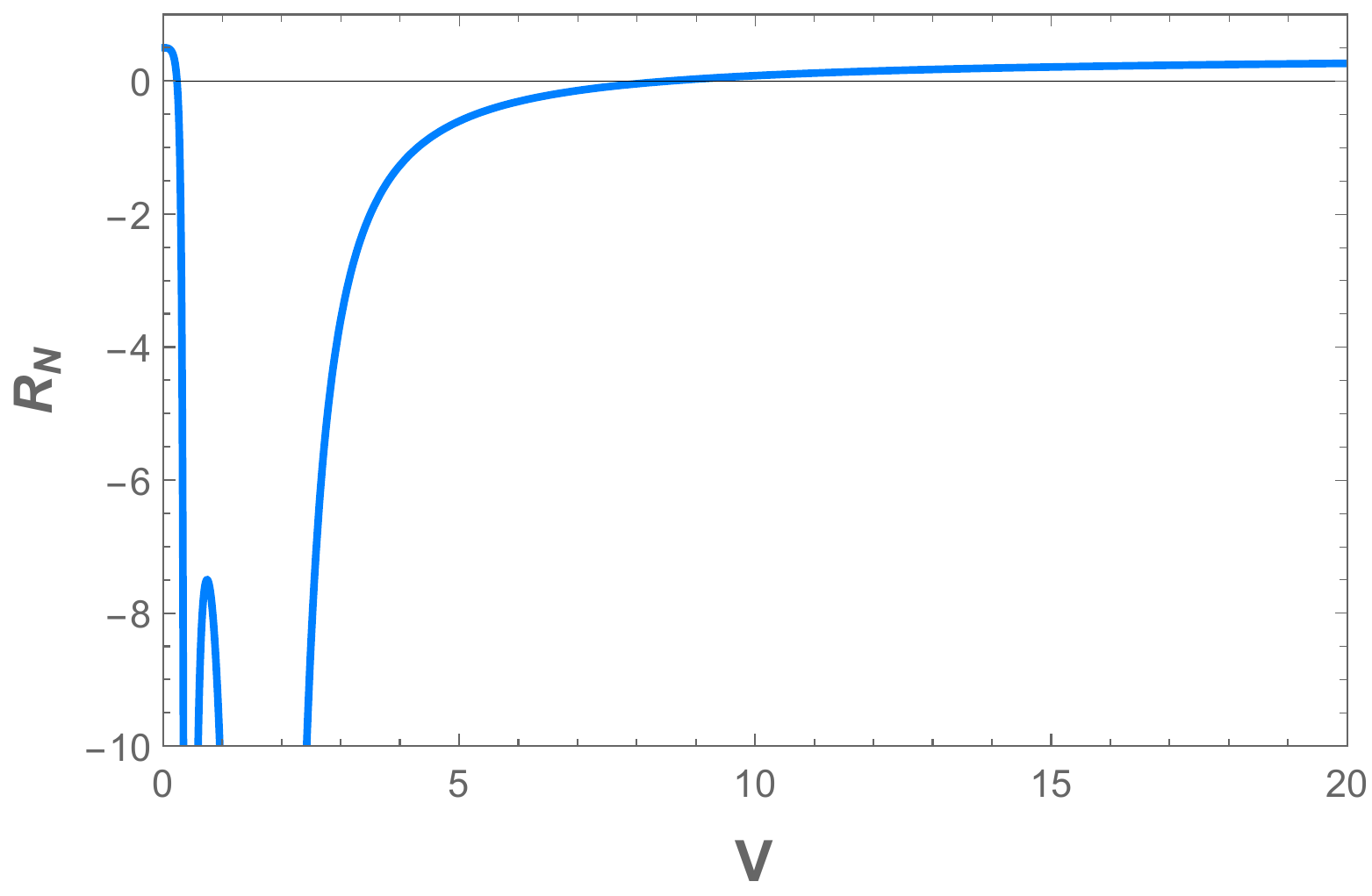}}\hspace{0.8cm}				
		\subfloat[]{\includegraphics[width=2.75in]{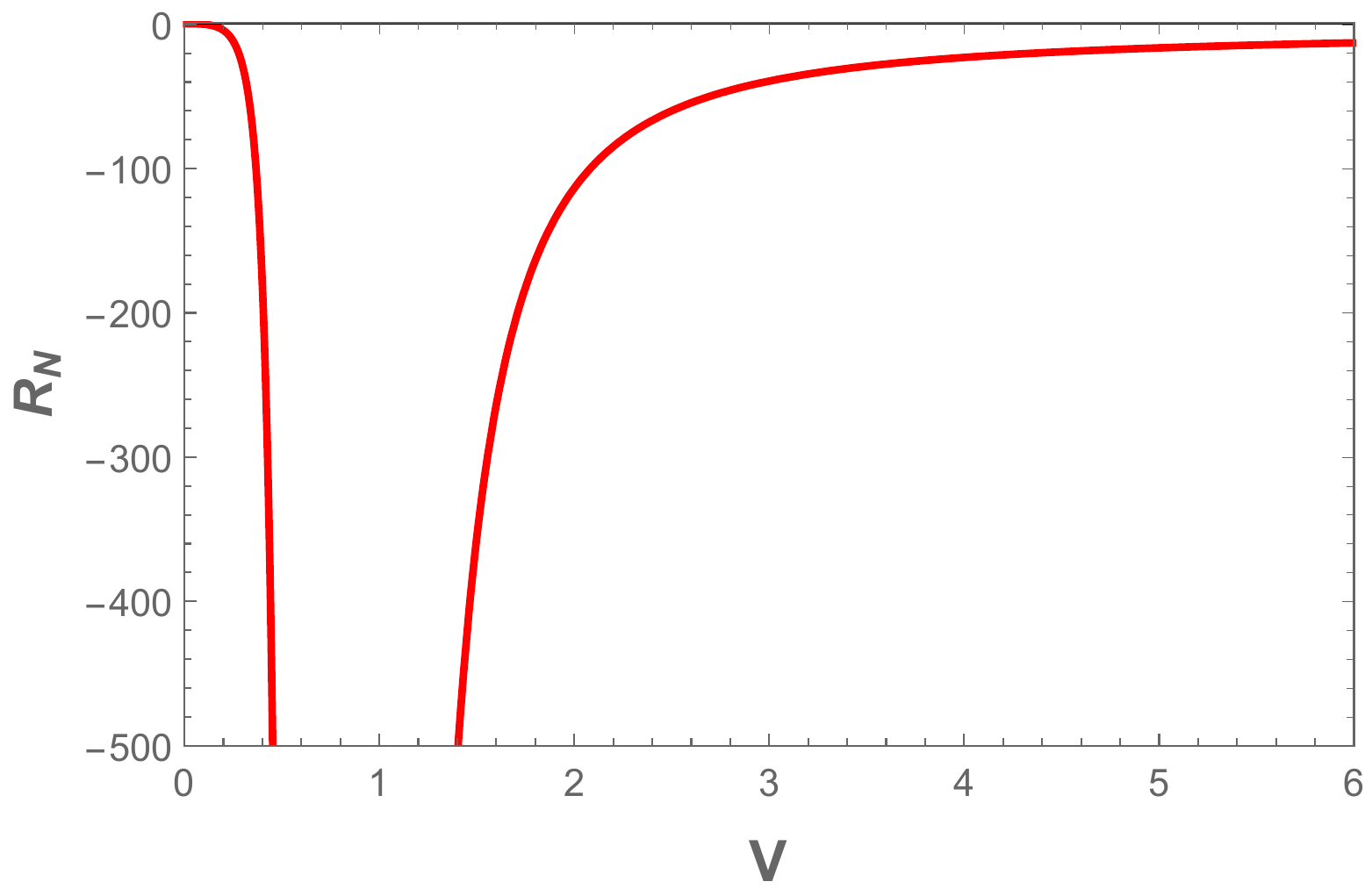}} \hspace{0.8cm}		
		\subfloat[]{\includegraphics[width=2.75in]{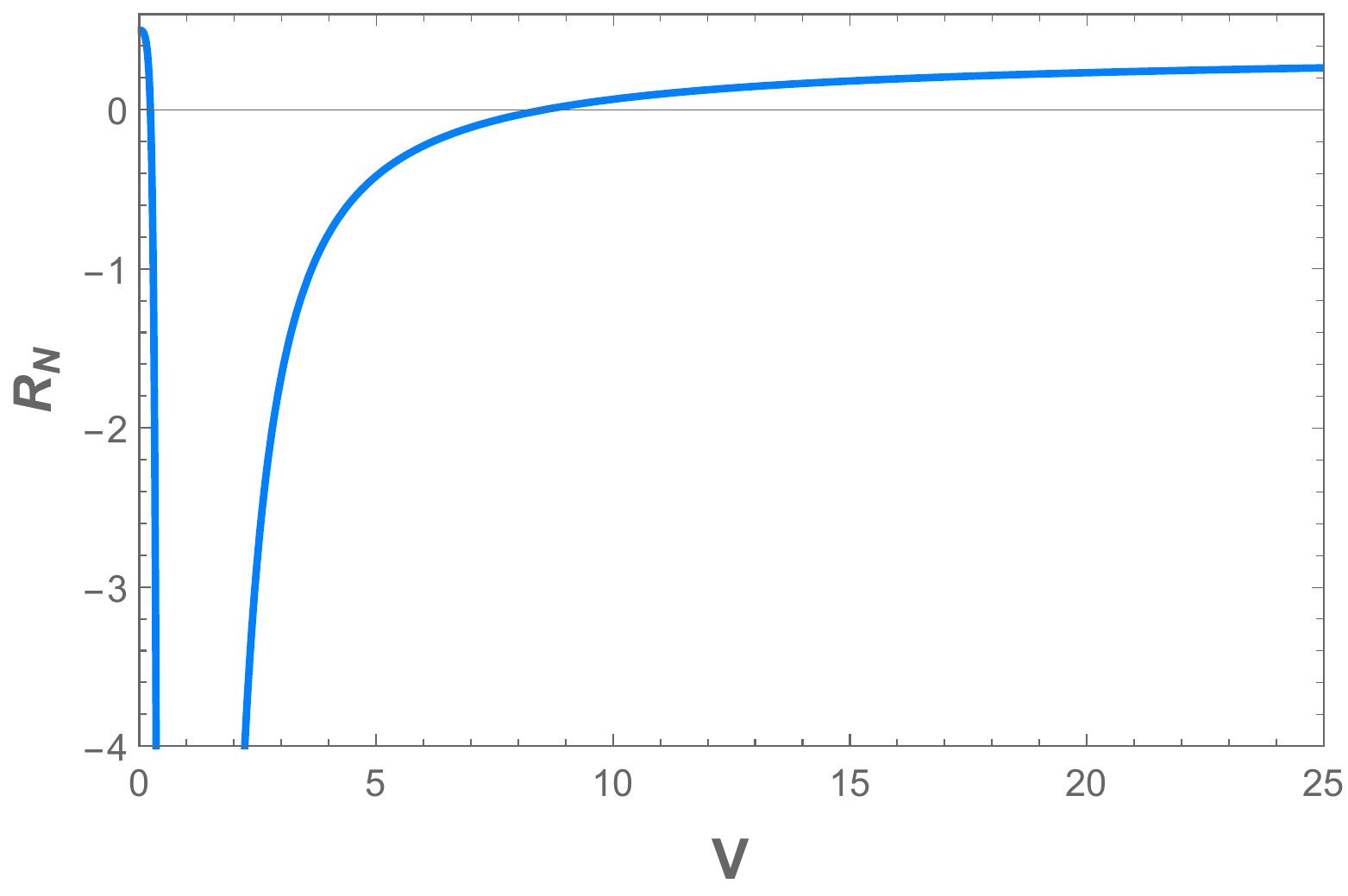}} \hspace{0.8cm}
		\subfloat[]{\includegraphics[width=2.75in]{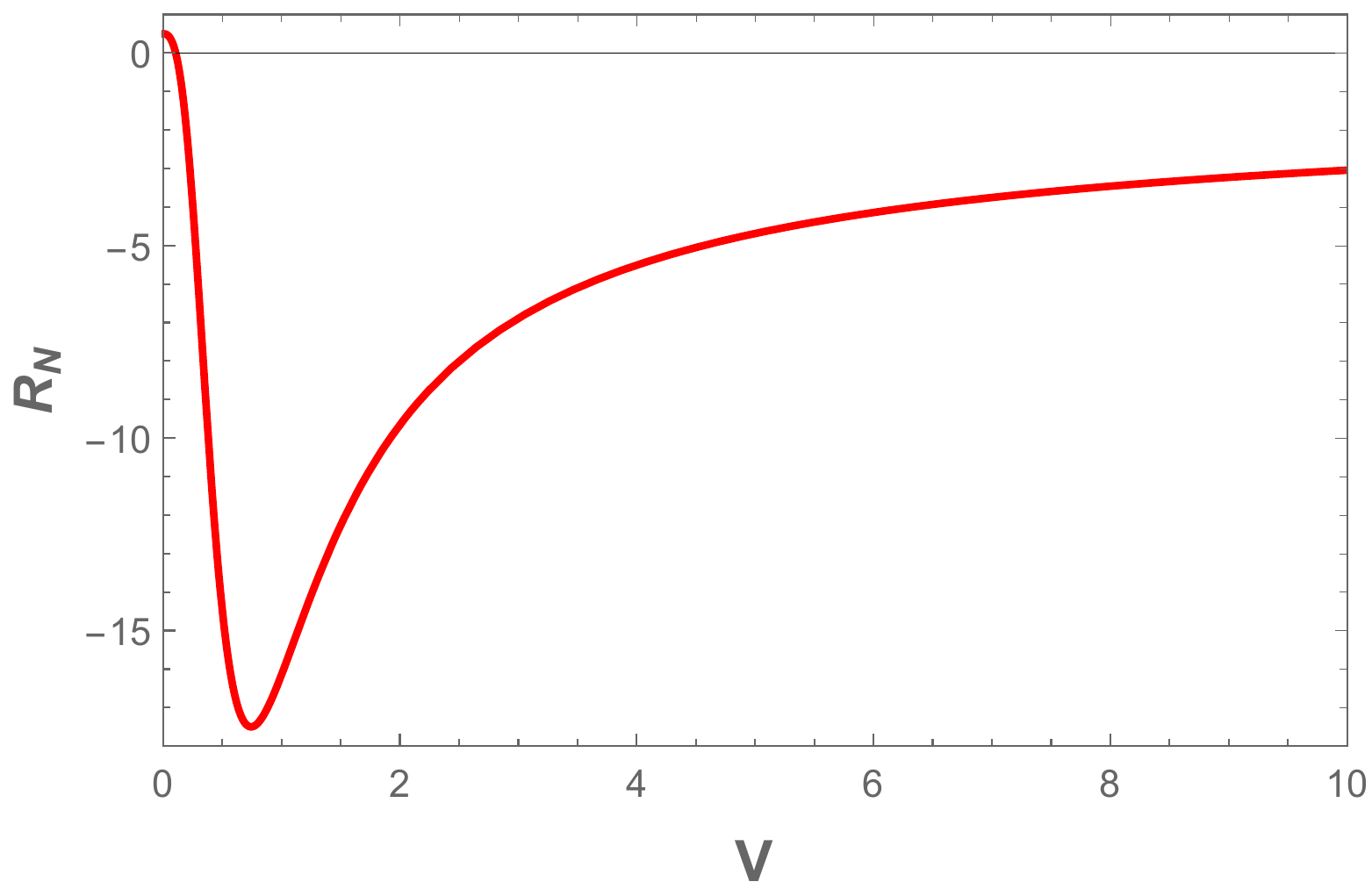}} \hspace{0.8cm}
		\subfloat[]{\includegraphics[width=2.75in]{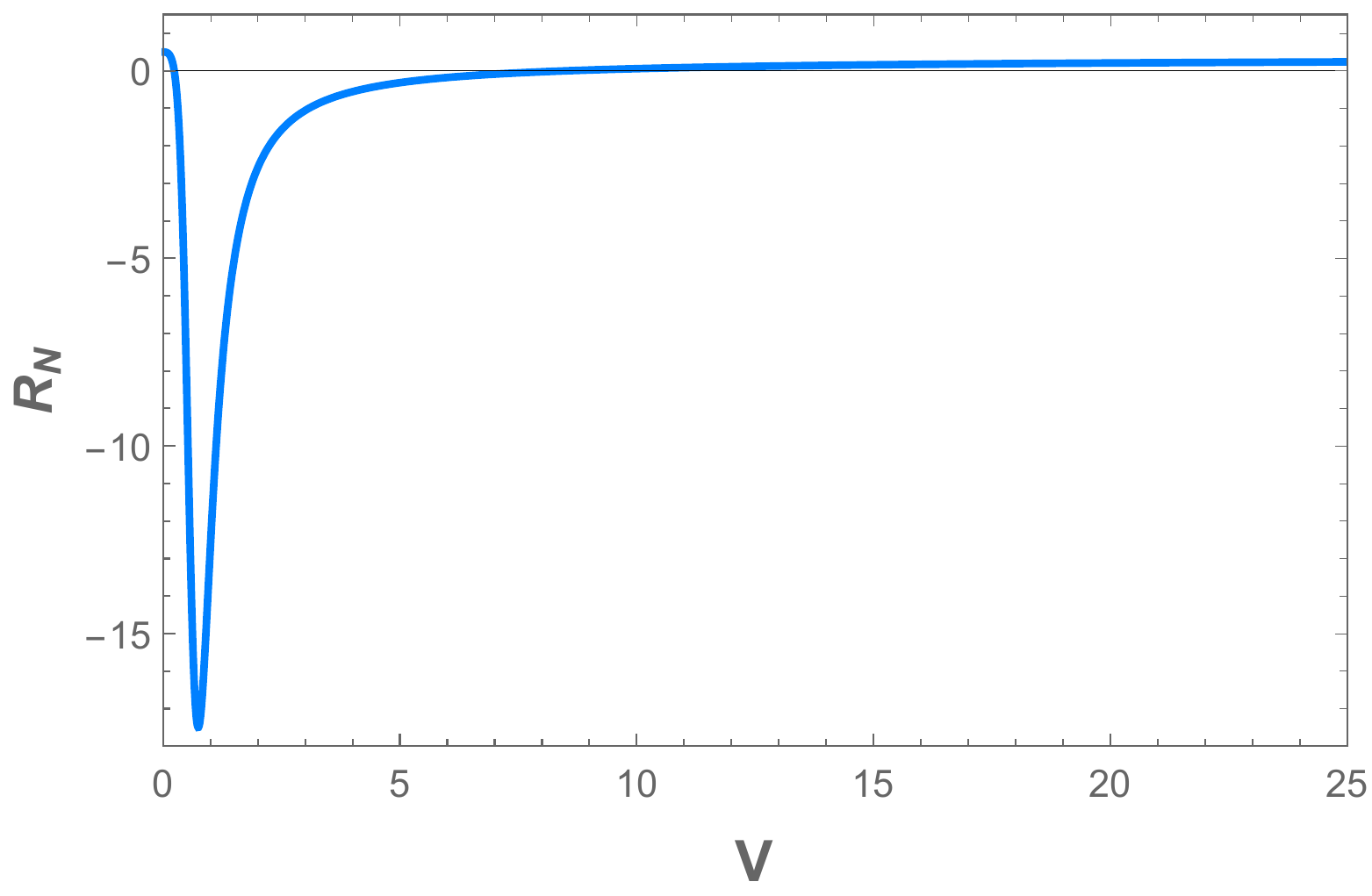}}

		\caption{Hyperbolic Topology case: \footnotesize   Behavior of $R_\text{N}$ with respect to $V$ at fixed $T$ : $c_1 = +7$ in left panel(red colored) and $c_1 = -7$ in right panel(blue colored).   (a)  $T = 0.495T_c$ (Inset shows the positive value of $R_\text{N}$ in some region of $V$). (b)  $T = 0.4T_c$.   $T =0.8T_c$ in (c) and (d). $T = T_c$ in (e) and (f).   $T =1.2T_c$ in (g) and (h).  
			(Here, we set the other parameters $ k=-1, \ q=1, \ m=2, \ c_0=1, \ c_2 =5.$)	}   
		\label{fig:RN_fixed_T_km1_plots}	}
\end{figure}

\noindent
{\underline { \bf Case-II: }} $c_1  \neq 0$:
\vskip 0.5cm
Next, we move to study the Ruppeiner geometry in the case of massive coefficient $c_1 \neq 0$, in non-reduced parameter space. In this case, the normalized scalar curvature $R_\text{N}$ for the equation of state~\eqref{eq of st}, turns out to be:
\begin{equation} \label{RNc1}
R_\text{N} =  \frac{\bigg( 2\epsilon(36\pi)^\frac{1}{3}V^\frac{2}{3}-16\pi q^2 + 3c_0 c_1 m^2V\bigg)\bigg(2\epsilon(36\pi)^\frac{1}{3}V^\frac{2}{3} -24\pi TV-16\pi q^2 + 3c_0 c_1 m^2V\bigg)}{2\bigg(2\epsilon(36\pi)^\frac{1}{3}V^\frac{2}{3} -12\pi TV-16\pi q^2 + 3c_0 c_1 m^2V\bigg)^2},
\end{equation} 
which diverges along the spinodal curve~\eqref{eq: spinodal curve c1 non zero}:
\begin{equation}
T_\text{SP} =  \frac{ 2\epsilon(36\pi)^{1/3}V^{2/3}-16\pi q^2 + 3c_0 c_1 m^2V}{12\pi V}.
\end{equation}
$R_\text{N}$ vanishes and changes the sign at the temperature $T_0 =\frac{T_\text{SP}}{2}$, and also when 
\begin{equation}\label{eq:cubic}
   2\epsilon(36\pi)^{1/3}V^{2/3}-16\pi q^2 + 3c_0 c_1 m^2V = 0.
\end{equation}
Provided the existence of criticality in any topology,  equation~\eqref{eq:cubic} gives one physical solution for $V$ when the massive  coefficient $c_1$ is positive, while two physical solutions  when  $c_1$ is negative.   
The implication of this result is that when the massive coefficient  $c_1$ is negative, the normalized scalar curvature $R_\text{N}$ has an additional sign changing curve as compared to the case of charged AdS black hole~\cite{Wei:2019uqg,Wei:2019yvs}.
\begin{figure}[h!]
	
	{\centering
		\subfloat[]{\includegraphics[width=3in]{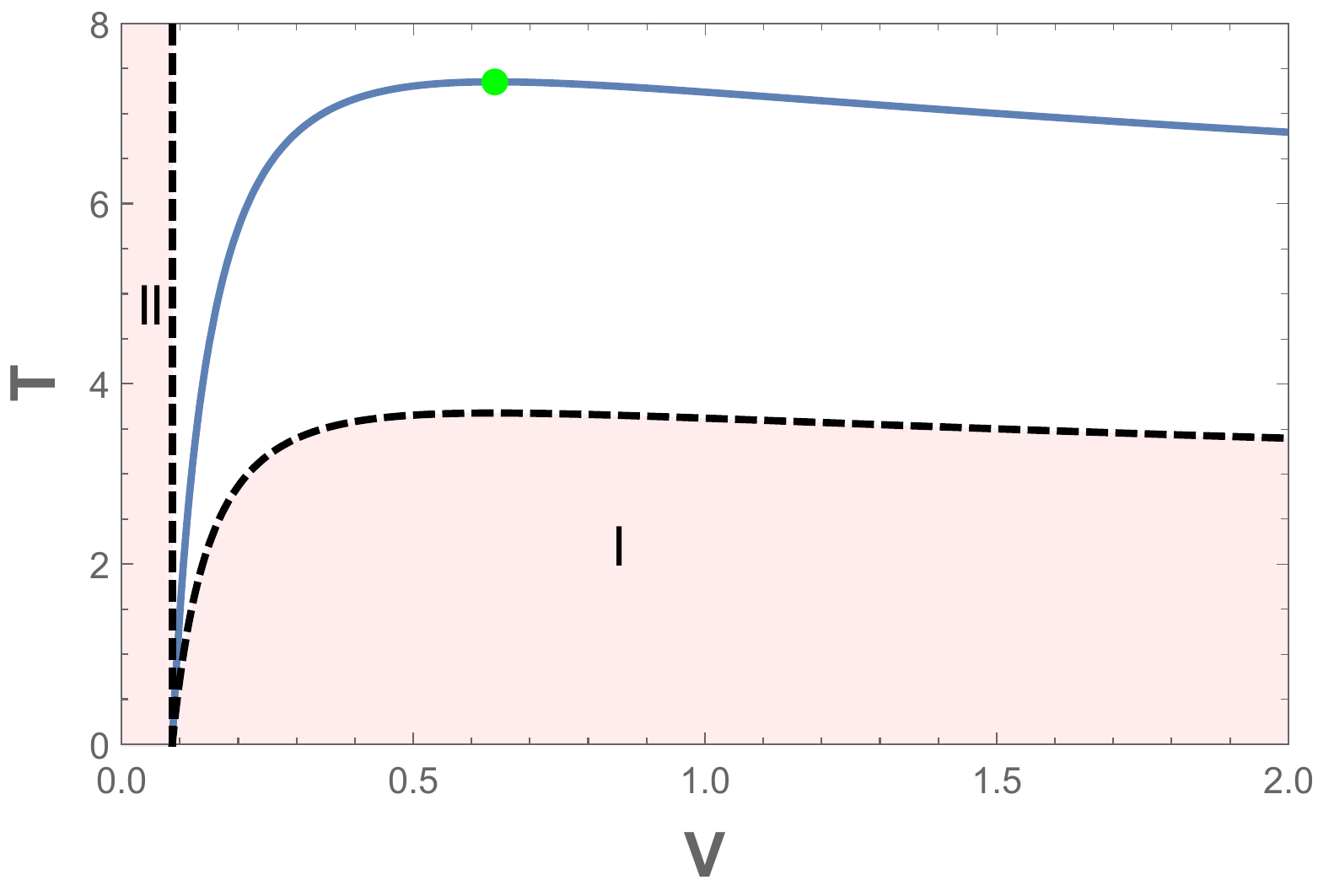}} \hspace{0.4cm}		
		\subfloat[]{\includegraphics[width=3in]{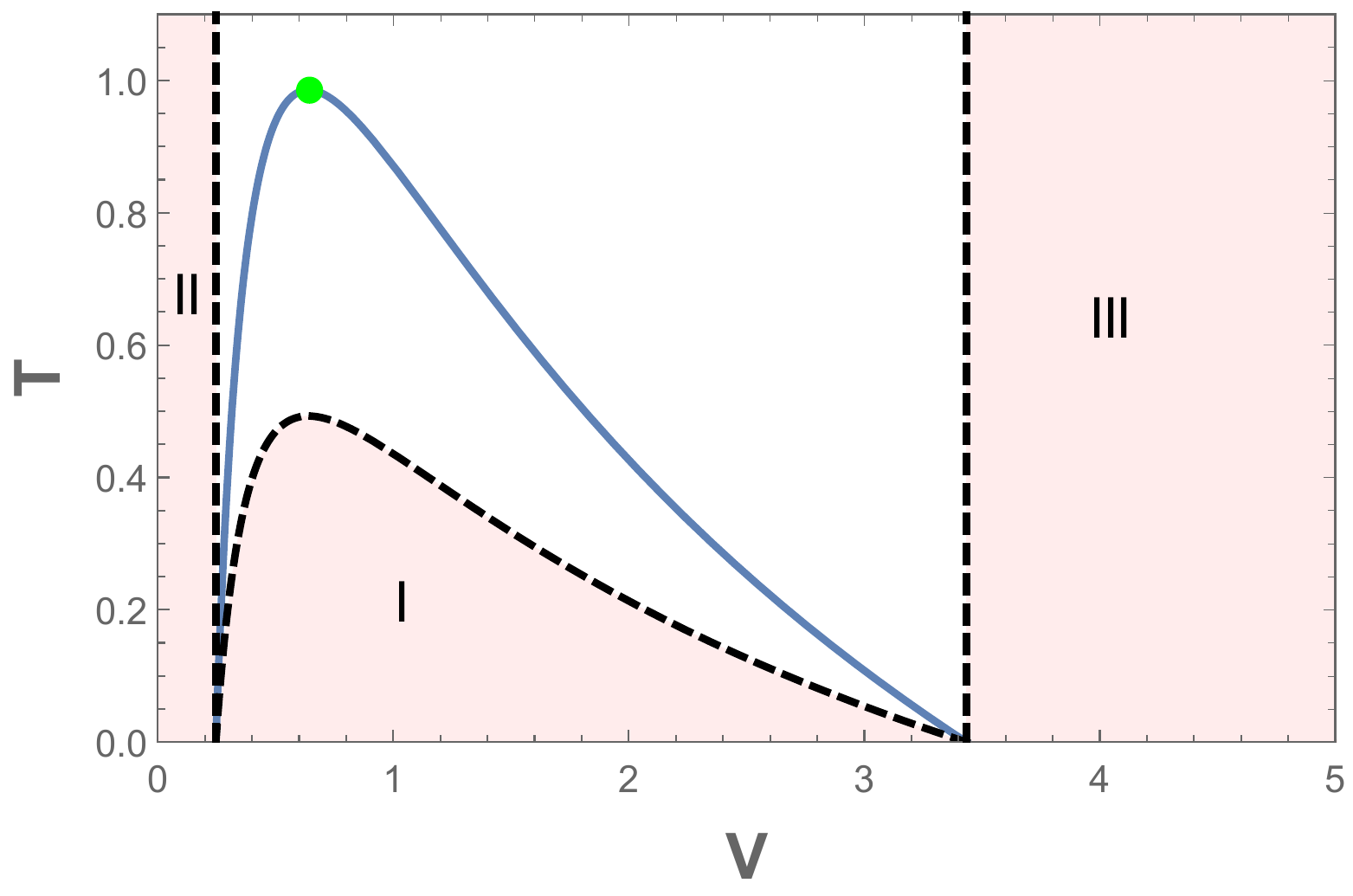}}

		\caption{Spherical Topology case: \footnotesize The spinodal curve (blue color), sign changing curves (dashed black color), and the critical point (green colored dot) are shown. (a) $c_1 = +10$. The spinodal curve and the sign-changing curves begin at $V=0.08687$. (b) $c_1 = -10$. The spinodal curve and the sign-changing curves begin at $V=0.24645$ and end at another sign changing curve at $V=3.43284$. $R_\text{N}$ is positive in the shadow regions, zero on dashed black curves(sign-changing curves), and negative in the remaining region.     
(Here, we set the other parameters $ k=1, \ q=1, \ m=2, \ c_0=1, \ c_2 =5.$)		}   
		\label{fig:tv_region_k1_plot}	}
\end{figure}
\begin{figure}[h!]
	
	{\centering
		\subfloat[]{\includegraphics[width=3in]{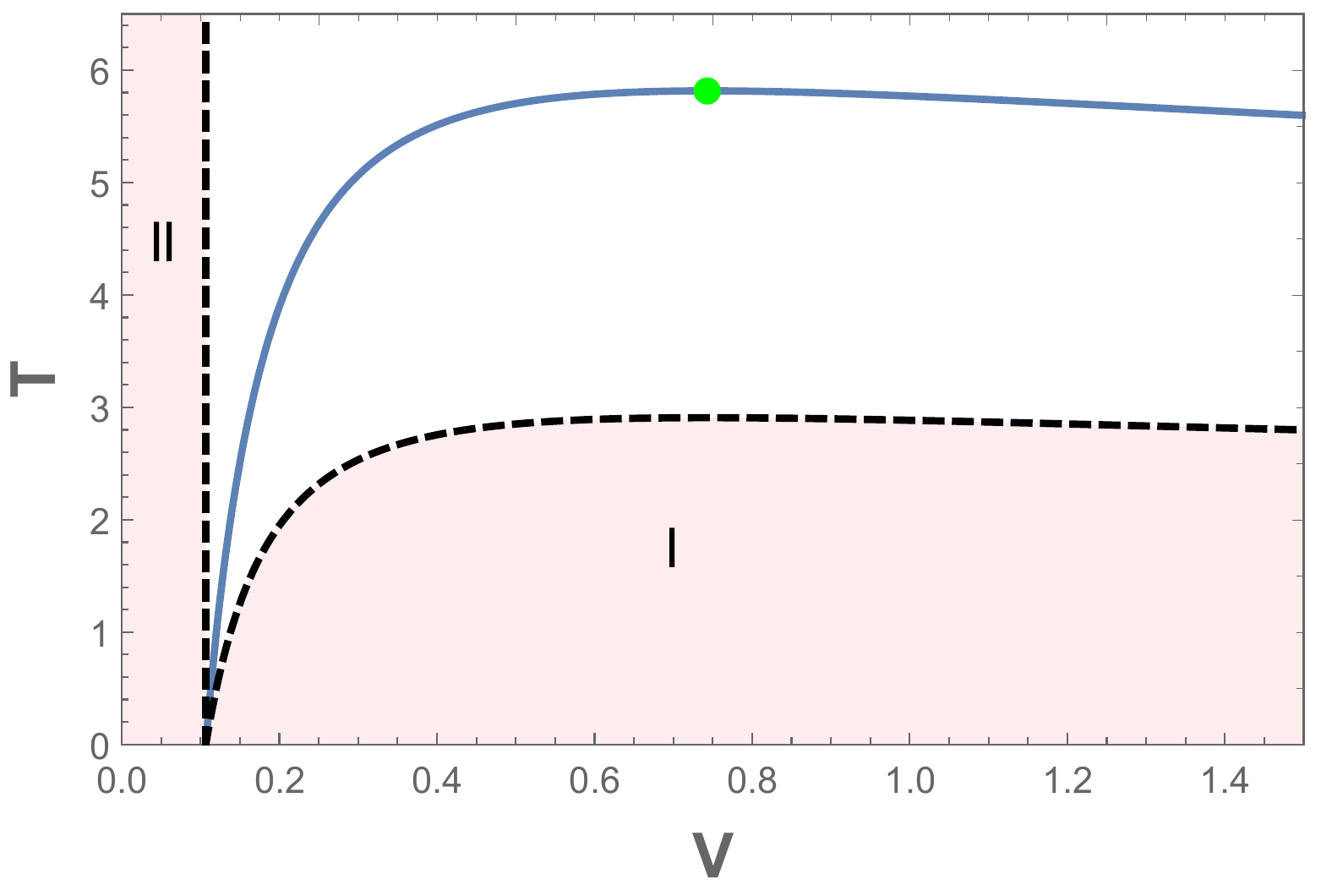}} \hspace{0.4cm}		
		\subfloat[]{\includegraphics[width=3in]{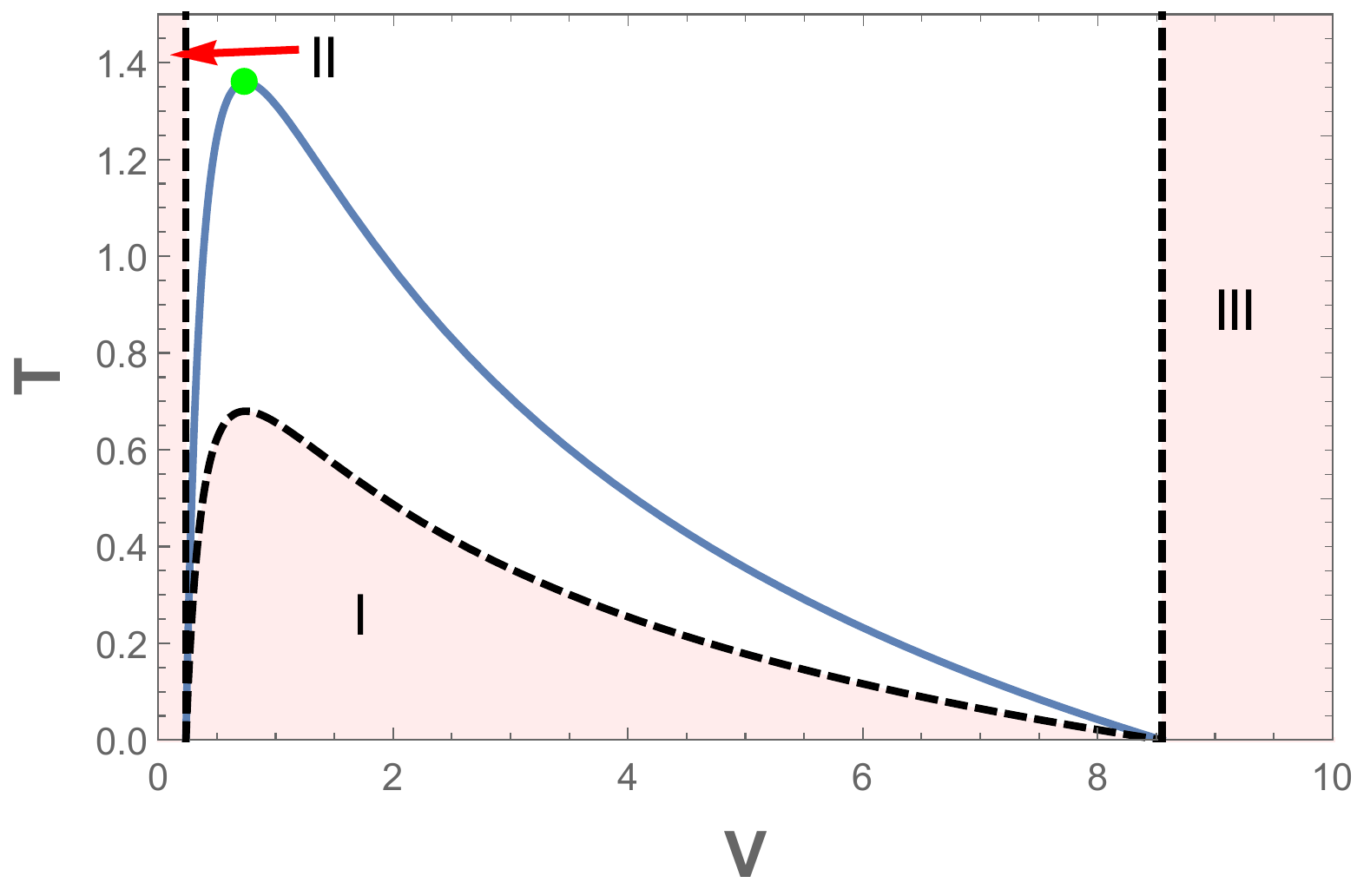}}

		\caption{Hyperbolic Topology case: \footnotesize The spinodal curve (blue color), sign changing curves (dashed black color), and the critical point (green colored dot) are shown. (a) $c_1 = +7$. The spinodal curve and the sign-changing curves begin at $V=0.10659$. (b) $c_1 = -7$. The spinodal curve and the sign-changing curves begin at $V=0.23522$ and end at another sign changing curve at $V=8.54663$. $R_\text{N}$ is positive in the shadow regions, zero on dashed black curves(sign-changing curves), and negative in the remaining region.     
			(Here, we set the other parameters $ k=-1, \ q=1, \ m=2, \ c_0=1, \ c_2 =5.$)		}   
		\label{fig:tv_region_km1_plot}	}
\end{figure}
\clearpage
We plot $R_\text{N}$ in Figures~\eqref{fig:3D_k1_plots} and~\eqref{fig:3D_km1_plots}, where it shows the divergence of $R_\text{N}$ along the spinodal curve which has the peak at critical point ($V_{c}$, $T_{c}$), and also below the critical temperature, $R_\text{N}$ has two diverging  points (one is at small $V$, and another is at large $V$).   
A more clear behavior of $R_\text{N}$ can be seen at fixed $T$ from Figures~\eqref{fig:RN_fixed_T_k1_plots} and~\eqref{fig:RN_fixed_T_km1_plots}, where we see that irrespective of topology, the two negative diverging points of $R_\text{N}$ come close as $T$ increases and become one at critical point;  and above the critical temperature $T_c$,  $R_\text{N}$ has an extremal point at $V_c$ without diverging.
Moreover, in addition to the presence of positive regions of $R_\text{N}$ at low temperatures, there exist a novel positive region of $R_\text{N}$ at larger thermodynamic volumes $V$ at any temperature when $c_1$ is negative. From Figures~\eqref{fig:tv_region_k1_plot} and~\eqref{fig:tv_region_km1_plot}, we can see the existence of this novel positive region-III of $R_\text{N}$ when the massive coefficient $c_1$ is negative. We see that region-III exists irrespective of the topology of the horizon. Despite the unavailability  of coexistence curve, which separates the small and large black holes, this novel positive region of $R_\text{N}$ always exist at larger thermodynamic volumes at any temperature and we can claim that the large black holes falling in this novel region exhibit the repulsive interactions among their microstructures.    

\section{Conclusions}\label{conclude}

In this paper, our aim was to  study, thermodynamics and phase transitions, including probing the interactions of microstructure of the  charged AdS black holes in massive gravity with various topologies of the horizon. Noting that the specific heat at constant volume $C_V=0$ for static black holes in this theory, we used the normalized scalar curvature $R_\text{N}$ of the Ruppeiner geometry~\cite{Wei:2019uqg,Wei:2019yvs},  assuming the temperature and thermodynamic volumes as fluctuating coordinates.
\vskip 0.3cm
\noindent
We observed that when the massive coefficient $c_1 = 0$, the scalar curvature $R_\text{N}$ depends on various parameters of the massive gravity system and can be shown to reduce to that of charged black holes in AdS, if one uses reduced variables at the critical point. The presence of massive graviton in Reissner-Nordstrom AdS black hole, makes the repulsive interactions of the microstructures of the small black hole  as strongly repulsive, while, the attractive interactions of the microstructures of the large black hole are weakly attractive.  Further, the repulsive interactions of the molecules of small black hole are strong for spherical topology $(k=+1)$, followed by flat topology $(k=0)$, and weak for hyperbolic topology $(k=-1)$, while, the attractive interactions of the molecules of the large black hole are strong for hyperbolic topology, followed by flat topology, and weak for spherical topology.
\vskip 0.3cm
\noindent
On the other hand, when the massive coefficient $c_1 \neq 0$: though the expression for coexistence curve of small and large black holes are unavailable\footnote{Although, one can naively define reduced variables as $,\tilde T = T/T_c,\tilde P = P/P_c, \tilde V = V/V_c$ using the expressions given in eqn. (\ref{eq:critical pt}), it is not useful as the resulting expressions will still depend on $c_1$ and charge $q$, unlike the Reissner-Nordstrom case considered in~\cite{Wei:2019uqg,Wei:2019yvs}.}, we argued that, provided the existence of criticality in any topology, the microstructure of the black holes is similar to that of the Reissner-Nordstrom AdS black holes when the massive coefficient $c_1$ is positive. However, when the   massive coefficient $c_1$ is negative, there exists a novel phase region for large black holes at any temperature. The implication of this novel phase region is that the large black holes falling in this region show repulsive interactions among their microstructures in contrast to the cases of  Reissner-Nordstrom AdS black hole and Van der Waals fluid~\cite{Wei:2019uqg,Wei:2019yvs}. It should be possible to confirm the above results using an alternate metric proposed recently, which gives a further clearer distinction of existence of attraction and repulsion dominated regions~\cite{Xu:2019gqm,Ghosh:2019pwy}, where the fluctuation coordinates can be either the combination $(T,V)$ or $(S,P)$. Furthermore, it would be interesting to obtain the mean field potential recently suggested in~\cite{Singh:2020tkf} to discuss the phase structure, and also see if it can specifically capture the new repulsion dominated regions found in this work for the massive gravity system.\\

\noindent
With regards to AdS/CFT correspondence, there have been various proposals to understand extended thermodynamics of black holes from the dual field theory side, known as black hole chemistry~\cite{Kubiznak:2016qmn}. For instance, varying the cosmological constant $\Lambda$ in the bulk corresponds to varying the number of colors $N$ in the boundary gauge theory, with chemical potential $\mu$ as its thermodynamic conjugate. In this context, studying Ruppeiner geometry of black holes in AdS with field theory variables is quite interesting, first pursued in~\cite{cai1} and recently revived in~\cite{Mahish:2020gwg}, by studying the microstructures giving important information. As this has not been explored, it would be very interesting to pursue this in the present example as well and see the effect of topology as well as graviton mass on the behavior of chemical potential and microstructures. In fact, the methods of holographic renormalization which have been developed in the case of massive gravity theories~\cite{Alishahiha:2010bw}, might prove to be useful.

\section*{Acknowledgements}
One of us (C.B.) would like to thank Aritra Ghosh for discussions on related topics.

\bibliographystyle{apsrev4-1}

\bibliography{massive_geometry}

\end{document}